\documentclass[sigconf,authorversion,pbalance=true]{acmart}
\AtBeginDocument{%
  }
\copyrightyear{2025}
\acmYear{2025}
\setcopyright{acmlicensed}\acmConference[ETRA '25]{2025 Symposium on Eye
Tracking Research and Applications}{May 26--29, 2025}{Tokyo, Japan}
\acmBooktitle{2025 Symposium on Eye Tracking Research and Applications
(ETRA '25), May 26--29, 2025, Tokyo, Japan}
\acmDOI{10.1145/3715669.3725872}
\acmISBN{979-8-4007-1487-0/2025/05}
\usepackage{enumitem}
\usepackage{amsmath}
\usepackage{fontawesome}

\usepackage{titlesec}
\usepackage{float}
\usepackage{subcaption}
\begin{document}
\title{Uncertainty-Aware Scarf Plots}

\author{Nelusa Pathmanathan}
\email{nelusa.pathmanathan@visus.uni-stuttgart.de}
\orcid{0000-0002-6848-8554}
\affiliation{%
\institution{University of Stuttgart}
  \city{Stuttgart}
  \country{Germany}
}
\author{Seyda Öney}
\email{seyda.\"oney@visus.uni-stuttgart.de}
\orcid{0000-0002-5785-6788}
\affiliation{%
\institution{University of Stuttgart}
   \city{Stuttgart}
  \country{Germany}
}

\author{Maurice Koch}
\email{maurice.koch@visus.uni-stuttgart.de}
 \orcid{0000-0003-0469-8971}
\affiliation{%
  \institution{University of Stuttgart}
   \city{Stuttgart}
  \country{Germany}
  }

\author{Daniel Weiskopf}
\email{daniel.weiskopf@visus.uni-stuttgart.de}
 \orcid{0000-0003-1174-1026}
\affiliation{%
 \institution{University of Stuttgart}
  \city{Stuttgart}
 \country{Germany}
 }
 
\author{Kuno Kurzhals}
\email{kuno.kurzhals@visus.uni-stuttgart.de}
\orcid{0000-0003-4919-4582}
\affiliation{%
  \institution{University of Stuttgart}
   \city{Stuttgart}
  \country{Germany}
}

\renewcommand{\shortauthors}{Pathmanathan et al.}

\begin{abstract}
Multiple challenges emerge when analyzing eye-tracking data with areas of interest (AOIs) because recordings are subject to different sources of uncertainties. Previous work often presents gaze data without considering those inaccuracies in the data. To address this issue, we developed uncertainty-aware scarf plot visualizations that aim to make analysts aware of uncertainties with respect to the position-based mapping of gaze to AOIs and depth dependency in 3D scenes. Additionally, we also consider uncertainties in automatic AOI annotation. We showcase our approach in comparison to standard scarf plots in an augmented reality scenario. 
\end{abstract}

\ccsdesc[500]{Human-centered computing~Visualization techniques}

\keywords{Eye tracking, uncertainty visualization, scarf plots, augmented reality, visualization}

\begin{teaserfigure}
    \centering
    \begin{subfigure}[b]{0.24\textwidth}
      \includegraphics[width=\textwidth]{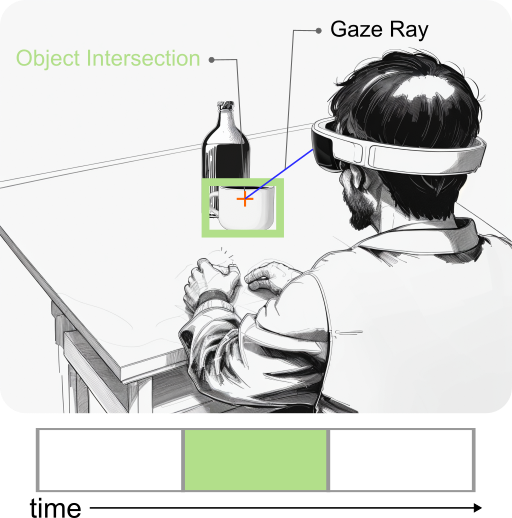}
      \caption{Standard scarf plot}\label{fig:algo1}
    \end{subfigure}
    \hfill
       \begin{subfigure}[b]{0.24\textwidth}
      \includegraphics[width=\textwidth]{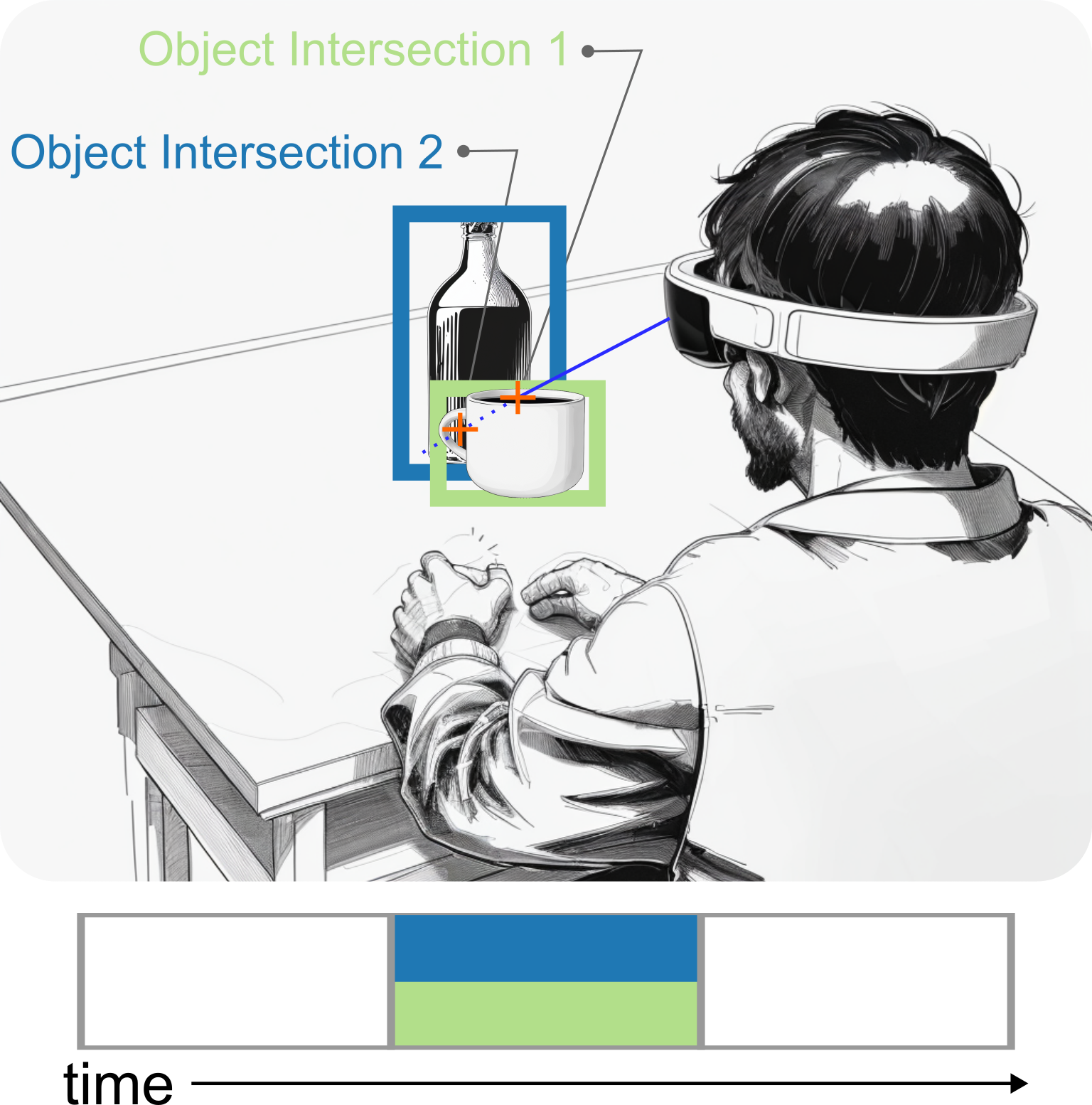}
      \caption{Depth scarf plot}\label{fig:algo2}
    \end{subfigure}
    \hfill
        \begin{subfigure}[b]{0.24\textwidth}
      \includegraphics[width=\textwidth]{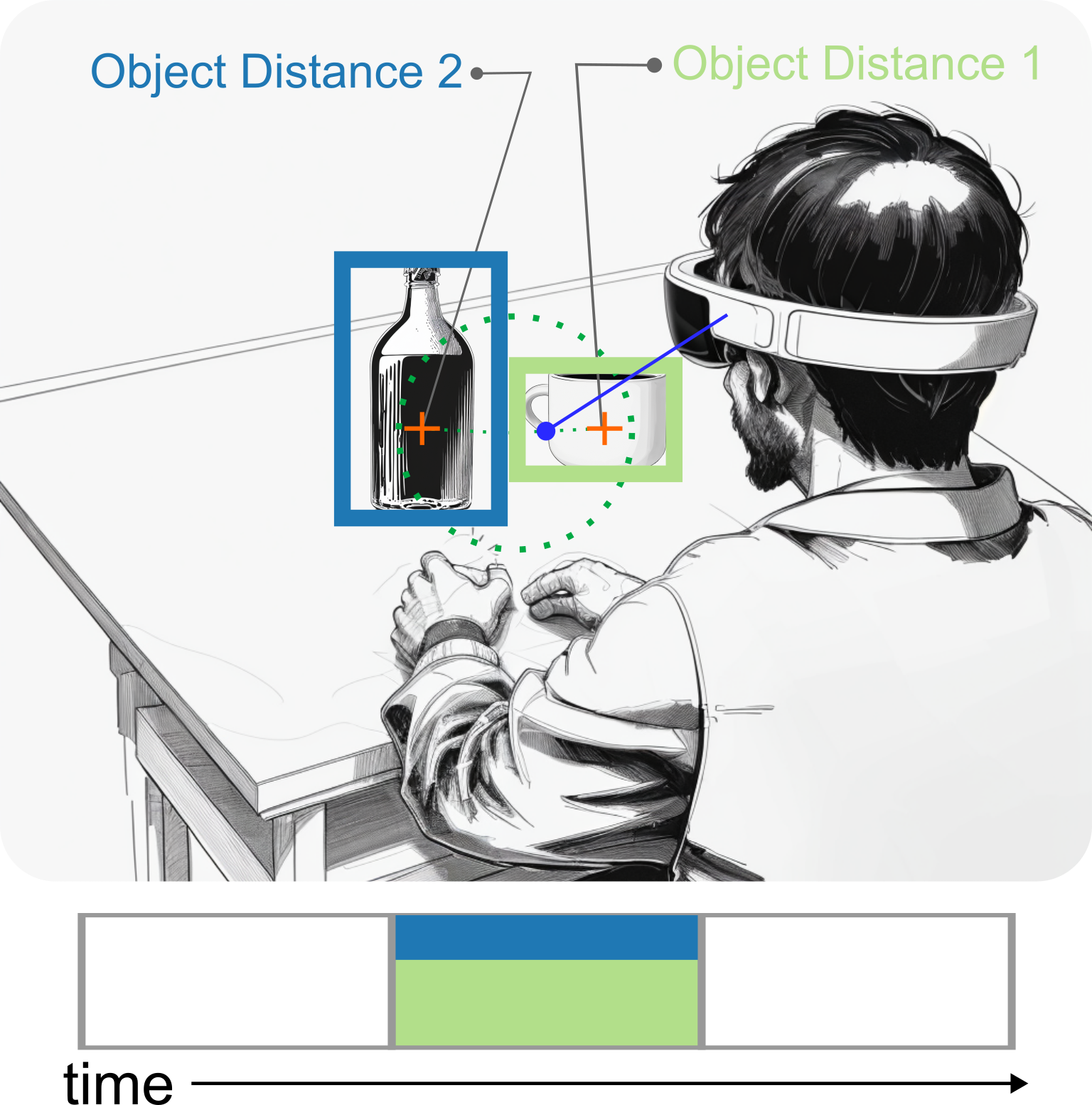}
      \caption{NN scarf plot}\label{fig:algo3}
    \end{subfigure}
    \Description{Three different images (a), (b) and (c) are depicted. In the first picture, a user with a head-mounted display (HMD) is sitting on a table with a cup in front of him and a bottle behind the cup. A line (gaze ray) starting in front of the person's head close to his eye area is drawn towards the cup, to show that he is looking at it. The cup is surrounded by a green 2D rectangular-shaped frame. Below the image, a standard scarf plot is visualized, showing three rectangular-shaped segments arranged sequentially, whereby the segment in the center is colored green, and the other two are white. Below the scarf plot an x-axis is drawn representing the timeline. The structure of the other two images is similar, where (b) draws a blue 2D rectangular-shaped frame around the bottle too. The green frame around the cup is annotated by the text "Object intersection 1," and the blue frame around the bottle is annotated by "Object intersection 2". To represent that both objects are getting intersected by the gaze ray, two red crosses are placed on the area where the line of the gaze ray intersects the cup and the bottle. The depth scarf plot below this figure looks similar to the standard scarf plot; however, this time, the middle segment is divided into two small sub-segments, one colored green, and the other blue. The last image (c) places the cup and the bottle next to each other, such that they are not occluded and are at the same depth on the table. Again both objects are surrounded by a 2D frame and annotated by a text saying "Object distance 1" for the green frame and "object distance 2" for the blue frame. On the center of the objects, a red cross is drawn. A dashed 2D circle is drawn, passing through the red crosses. This time, the gaze ray does not intersect any of the objects and ends in between both objects (a little bit closer to the cup than the bottle).}
    \caption{
(a)~\textit{Standard scarf plots} are used to analyze gaze patterns on AOIs over time. These consider direct hits with single AOIs, represented by one segment (e.g., green segment for hit with cup AOI). 
(b)~\textit{Depth scarf plots} consider gaze-ray intersections with multiple AOIs to account for uncertainties due to AOI overlaps and depth order (first cup (green), second bottle (blue)). 
(c)~\textit{Nearest neighbor (NN) scarf plots} account for AOIs near the gaze ray.
The height of a sub-segment is proportional to the distance between AOI and gaze ray
(e.g., cup (green) is closer to the gaze ray than bottle (blue)). (Images edited with Stable~Diffusion)
    }
    \label{fig:algorithm}
\end{teaserfigure}

\maketitle
\section{Introduction}
Eye-tracking data is often quantified and semantically enriched by areas of interest (AOIs)~\cite{holmqvist2011eye, raschke2014}. 
AOIs support the interpretation of scanpaths as a sequence of visited regions (e.g., different objects) and are the data basis for many statistical and visual analysis methods~\cite{blascheck2017taxonomy, holmqvist2011eye}.
The definition of AOIs in 2D is often implemented by bounding shapes, either by simple bounding boxes or complex polygonal shapes and a semantic label. 
The mapping of AOIs to individual gaze samples or fixations is then achieved by hit detections with the corresponding shapes. While this approach has a long tradition, it comes with some inherent problems considering inaccuracies on different levels: 

\begin{enumerate}[leftmargin=*]
\item \textit{Gaze is not a point.} The foveated area covers an angle of about 2 degrees~\cite{strasburger2011peripheral, KosslynVisualAngle}, allowing for potentially multiple hits with different AOIs simultaneously. This effect is amplified by inaccuracies of the eye-tracking device.

\item \textit{Bounding shapes are often not optimal.} Simple bounding boxes often include more scene content than just the AOI. Pixel-perfect shapes are often used in computer vision applications as ground-truth data but require much more annotation effort and still face the problem of inaccuracies in measurements.

\item \textit{Overlaps of AOIs happen.} In dynamic scenes, AOIs will often overlap, leading to ambiguities in the mapping step. 

\item \textit{Misclassifications of AOIs.} Computer vision \cite{de2012automatic, deane2023deep} supports the classification of AOIs. Such techniques also include inaccuracies that lead to false classifications or missing of AOIs.
\end{enumerate}
These issues further increase with 3D stimuli as they occur in eye-tracking scenarios in virtual reality (VR) or augmented reality (AR) \cite{kurzhals2022Situated, oney2023_AR}.
Shapes have to be extended to bounding volumes, and the additional degrees of freedom increase the effort for precise annotation immensely.
While it is worthwhile reducing inaccuracies in all of these aspects, some uncertainty in the data will always remain.
Hence, our main research question is: 

\textit{How can we make analysts aware of the inaccuracies in gaze data and automated AOI detection?}    

To address this issue, we suggest a visualization that extends an established technique for AOI-annotated scanpath sequences: the \textit{scarf plot} \cite{richardson2005looking}.
In summary, our contributions are:
(1) A new technique, \textit{uncertainty-aware scarf plots}, that extends the concept of scarf plots to visualize inherent uncertainty in gaze data and object recognition in stimuli.
(2) Demonstrating our approach within an AR scene consisting of physical and virtual objects. The physical objects are defined as AOIs with automatic detection. We show how our approach depicts ambiguities in hit detection (uncertainty in position), object detection (uncertainty in classification), and depth (uncertainty in order) differences in the~data.

\section{Related Work}
The AOI-based analysis of gaze data requires the definition of AOIs, which is often time-consuming and can be supported by semi-automatic labeling~\cite{Kurzhals2017Clustering,pontillo2010, barz2023interactive, kopacsi2023imeta} and automatic classification~\cite{wolf2018automating, panetta2019software,deane2023deep}. In our approach, we also employ an object recognition algorithm that allows for automatic detection and labeling of AOIs.

There are various visualization techniques that facilitate AOI analysis~\cite{blascheck2017taxonomy}. 
\citet{richardson2005looking} introduced the scarf plot visualization, where
gaze data are aggregated into dwell times and color-coded according to labeled AOIs. 
Variations of scarf plots have been developed~\cite{falzone2023using, kurzhals2014iseecube, andrienko2012visual, weibel2012let, yang2020comparison,andrienko2022seeking}. \textit{Alpscarf} by  \citet{Yang2018alpscarfs} extends scarf plots vertically to visualize patterns of sequential conformity and gaze revisits. 
\citet{wang2021object,wang2023gaze} incorporated peripheral vision into the AOI assignment by computing the distance between the point of regard to nearby AOIs (object-to-gaze distance).
We employ a similar concept based on inverse distance weighting~\cite{Shepard1968} to define uncertainty caused by positional offsets.
The integration of eye tracking into Head-Mounted Displays (HMDs) has further enabled gaze-based analysis in AR studies, opening up new possibilities for evaluation~\cite{kurzhals2022Situated,dzsotjan2021predictive}.
Spatial context facilitates the automatic labeling of virtual AOIs~\cite{pfeiffer2014eyesee3d,oney2023_AR}. 
Our approach uses the spatial context in ambiguous cases by considering the distance of each AOI from the gaze position and visualizing all potential AOIs with their respective probabilities of spatial proximity to the gaze ray.

There are many different visualization techniques to highlight data uncertainty in general~\cite{Bonneau:2014:OSU,Griethe:2006:VUD,Padilla:2020:UV,weiskopf2022uncertainty}.
A specific approach for eye-tracking data \cite{Koch2024ActiveGaze} focuses on uncertainties from misclassifications of AOIs in automated labeling methods, enabling manual correction of automatic classifications and increasing user trust in automated methods.
\citet{Wang2022GazeUncertainty} investigated the impact of gaze uncertainty on information visualizations.  
They defined uncertainty on a fixation level as the extent of spatial overlap between gaze samples and AOIs, and metrics to quantify the impact of gaze uncertainty.  
In our work, we consider both inaccuracies in AOI recognition and uncertainties in gaze data. Addressing both aspects enables a deeper analysis of uncertainties at the level of both AOI and gaze data, which improves the interpretability of eye-tracking data.

\section{Technique}
\begin{figure*}[!htbp]
  \centering  
  \begin{subfigure}[b]{0.3\textwidth}
      \includegraphics[width=\textwidth]{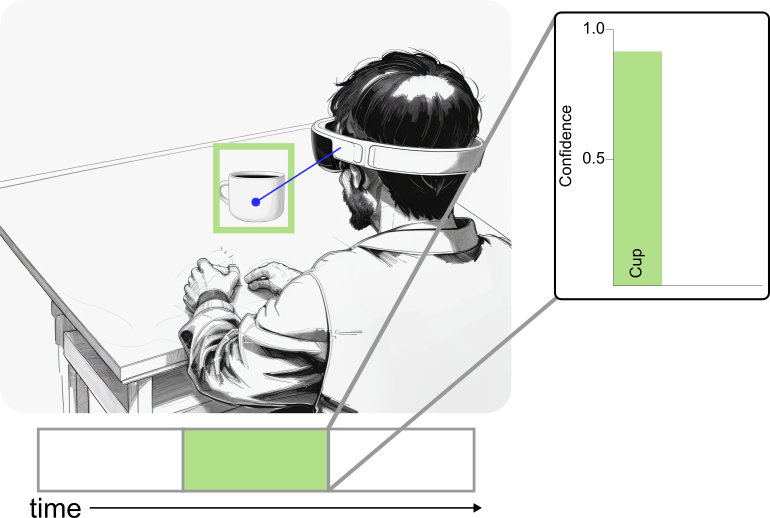}
      \caption{True Positive (TP)} \label{fig:fig:algo-4-5}
    \end{subfigure} 
    \hfill%
    \begin{subfigure}[b]{0.3\textwidth}
      \includegraphics[width=\textwidth]{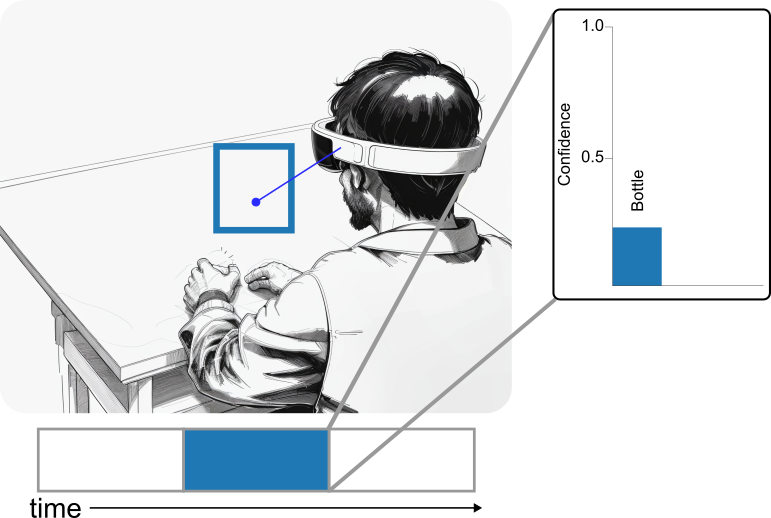}
      \caption{False Positive (FP) 1}\label{fig:algo-4-1}
    \end{subfigure}
    \hfill%
       \begin{subfigure}[b]{0.3\textwidth}
      \includegraphics[width=\textwidth]{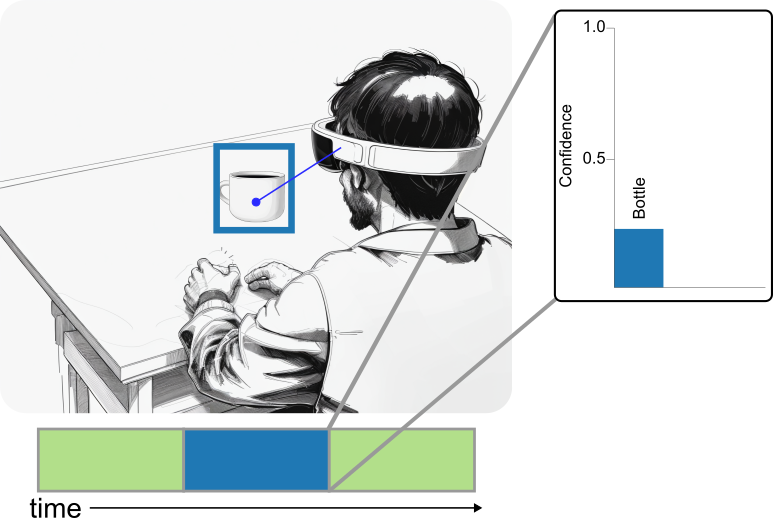}
      \caption{FP 2}\label{fig:fig:algo-4-2}
    \end{subfigure}
    \Description{
      Three different images (a), (b), and (c) are depicted. In the first picture, a user with a head-mounted display (HMD) is sitting on a table with a cup in front of him. A line (gaze ray) starting in front of the person's head close to his eye area is drawn towards the cup, to illustrate that he is looking at it. The cup is surrounded by a 2D green frame. Below the image, a standard scarf plot is visualized, showing three rectangular-shaped segments arranged sequentially, whereby the segment in the center is colored green, and the other two are white. Below the scarf plot an x-axis is drawn representing the timeline. Besides the image a bar chart is depicted with one bar in green, representing the confidence of detection for the cup. In (b), there is no object on the table but a blue 2D frame on which the person is looking; the segment in the center of the scarf plot is blue, and the bar chart shows a blue bar with a length below 0.5. The last figure (c) shows a cup, which is surrounded by a blue frame, the segment in the middle of the scarf plot is blue, while the other two are green. The barchart is similar to the one in (b).
    }
    \caption{(a) shows a True Positive (TP) classification, where the cup is classified correctly with high confidence. There can be two types of false positive (FP) classification. (b) There is no object, but a bottle gets detected, with low confidence. (c) A cup is detected as bottle, with low confidence. (Images edited with Stable Diffusion)
    }\label{fig:algo4}
\end{figure*}

\subsection{Types of Uncertainty}
\label{sec:typesuncertainty}
\paragraph{Uncertainty in position}
One common cause for uncertainty are positional offsets,
often compensated by defining additional margins around AOI bounding boxes. 
In some cases, a gaze ray cannot be uniquely assigned to a single AOI even with compensatory margins. When a gaze ray falls within one bounding box but is close to a neighboring AOI, determining the correct assignment becomes challenging. This also applies when the gaze ray does not intersect any bounding box but lies between two AOIs. Thus, it is important to indicate that the gaze may correspond to both AOIs, although one may be more likely than the other. 
\paragraph{Uncertainty in order}
In 2D and 3D, overlaps of AOIs can cause ambiguities for hit detections. 
While close objects are often preferred as a first hit (\autoref{fig:algo1}), partial occlusion and transparency are potential causes for ambiguous mappings. These overlaps often occur when objects are positioned along a line at different depths. 
In AR, this scenario occurs, for example, when a virtual object is placed in front of another object. 
In this case, the encoding of depth informs the analyst about the object in the background and the possibility that the user looked at it.
\paragraph{Uncertainty in classification}
Considering the confidence of a classifier becomes important if automated procedures are applied for AOI detection.
Depending on the classifier, the results can lead to an automated annotation with misclassified objects. It is important to provide information regarding the confidence in an AOI assignment. This information can help analysts reason about inconsistencies within the visual flow of a scarf plot. ~\autoref{fig:algo4} shows the different classification errors. A bar chart linked to the scarf plot shows the percentage of confidence for a misclassified object.

\subsection{Scarf Plot and Its Extension to Uncertainty}
\label{sec:scarfplotextension}
A scarf plot \cite{richardson2005looking} is 
a temporal alignment of segments (e.g., frames, gaze samples, fixations) encoded by color, which depicts the AOI the segment was assigned to (\autoref{fig:algorithm}), revealing a pattern evolving over time. This allows us to answer questions such as which AOIs received the most attention, the order in which the AOIs were visited, and the time spent viewing a specific AOI. 
Common techniques assign the segment to the AOI that was hit first by the point of regard. 
By extending the design of the scarf plot, we want to reveal the previously hidden uncertainties discussed above, 
to achieve more reliable results during analysis.

\begin{figure*}[t]
  \centering
       \begin{subfigure}[b]{0.24\textwidth}
      \includegraphics[width=\textwidth]{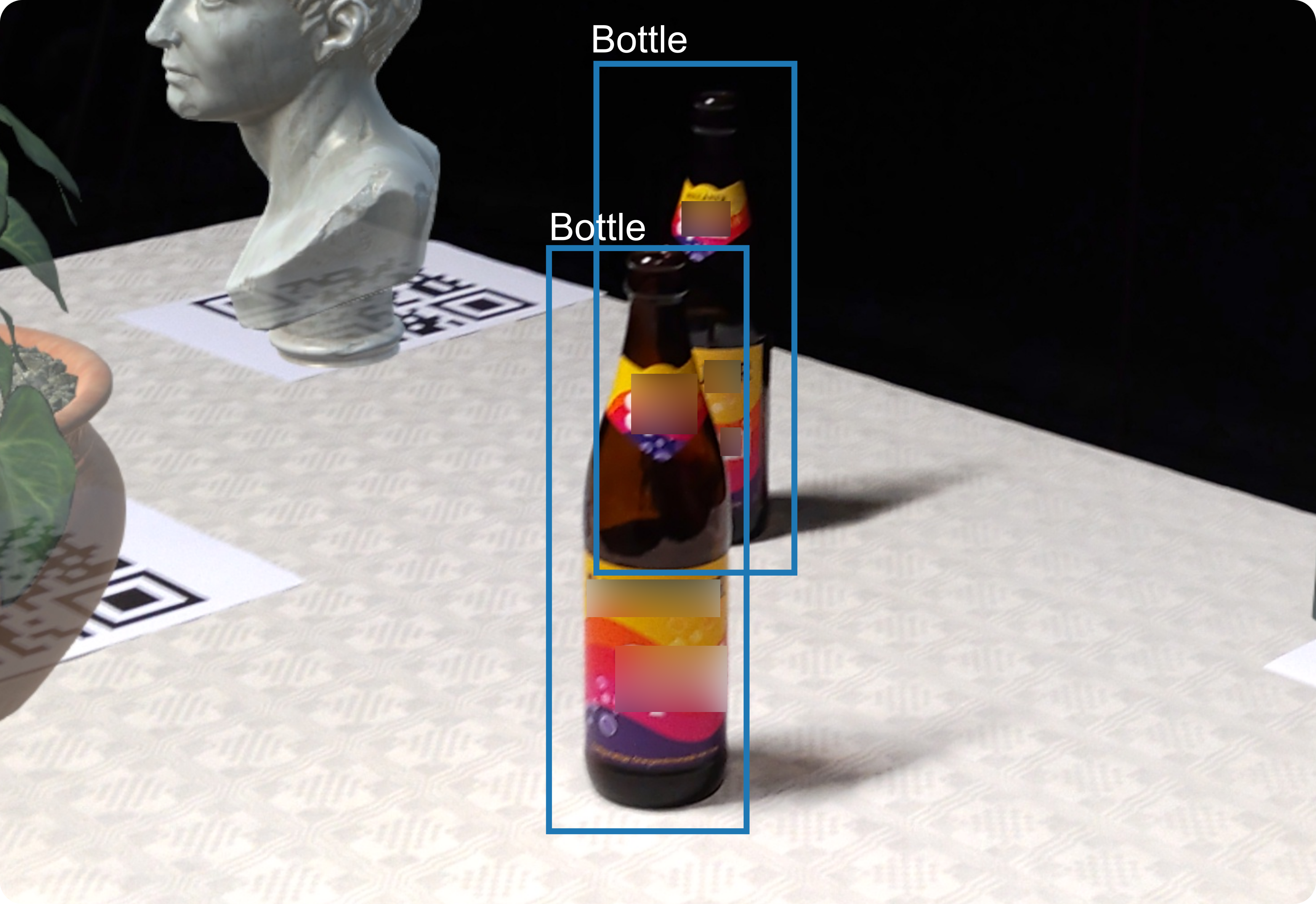}
      \caption{2 bottles (BB)\\ \mbox{}}\label{fig:2bottles}
    \end{subfigure}
        \begin{subfigure}[b]{0.24\textwidth}
      \includegraphics[width=\textwidth]{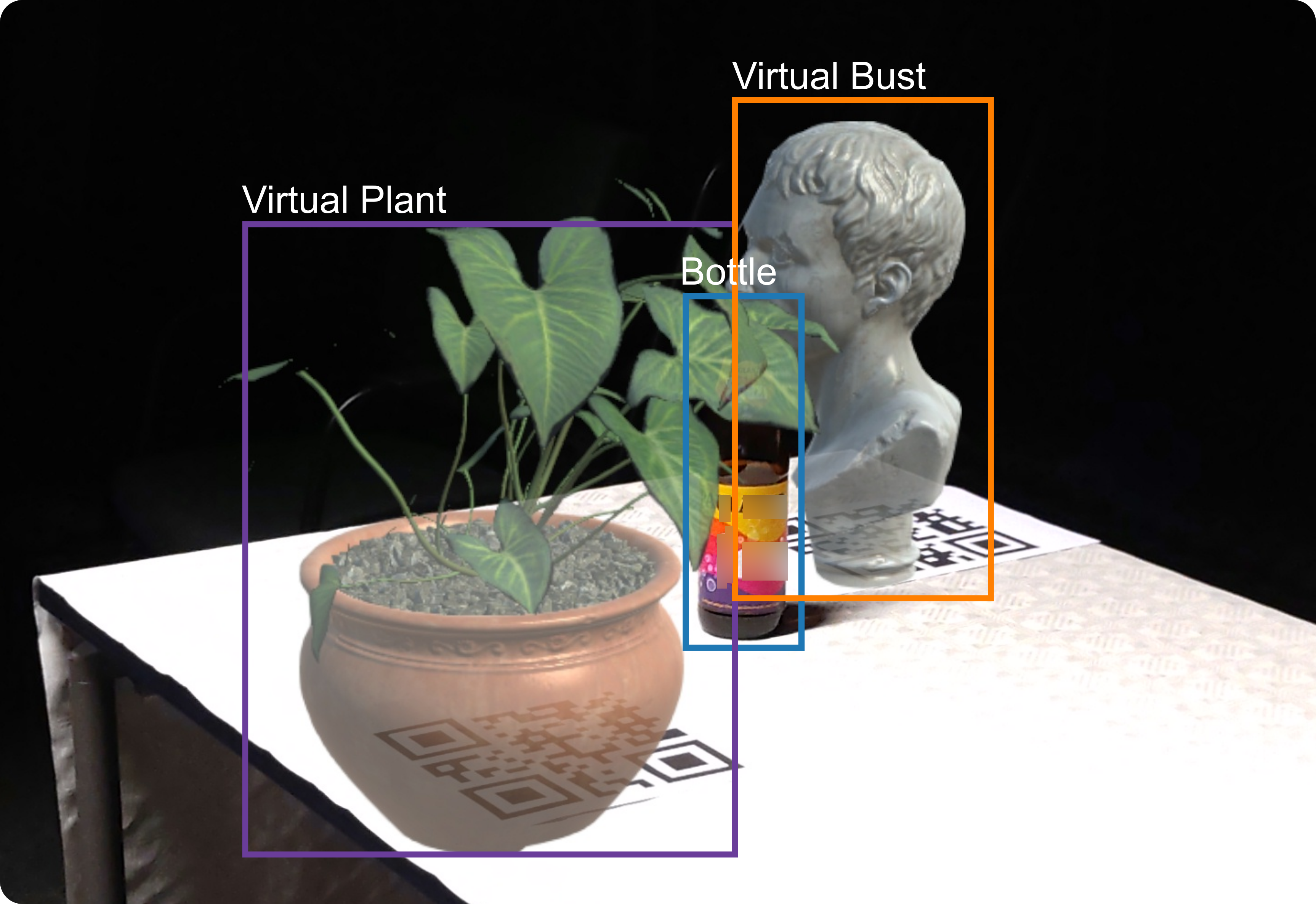}
      \caption{Virtual plant, bottle and \\
      virtual bust (VP, B \& VB)}\label{fig:virutalbottle}
    \end{subfigure}
    \vspace{3ex}
        \begin{subfigure}[b]{0.24\textwidth}
      \includegraphics[width=\textwidth]{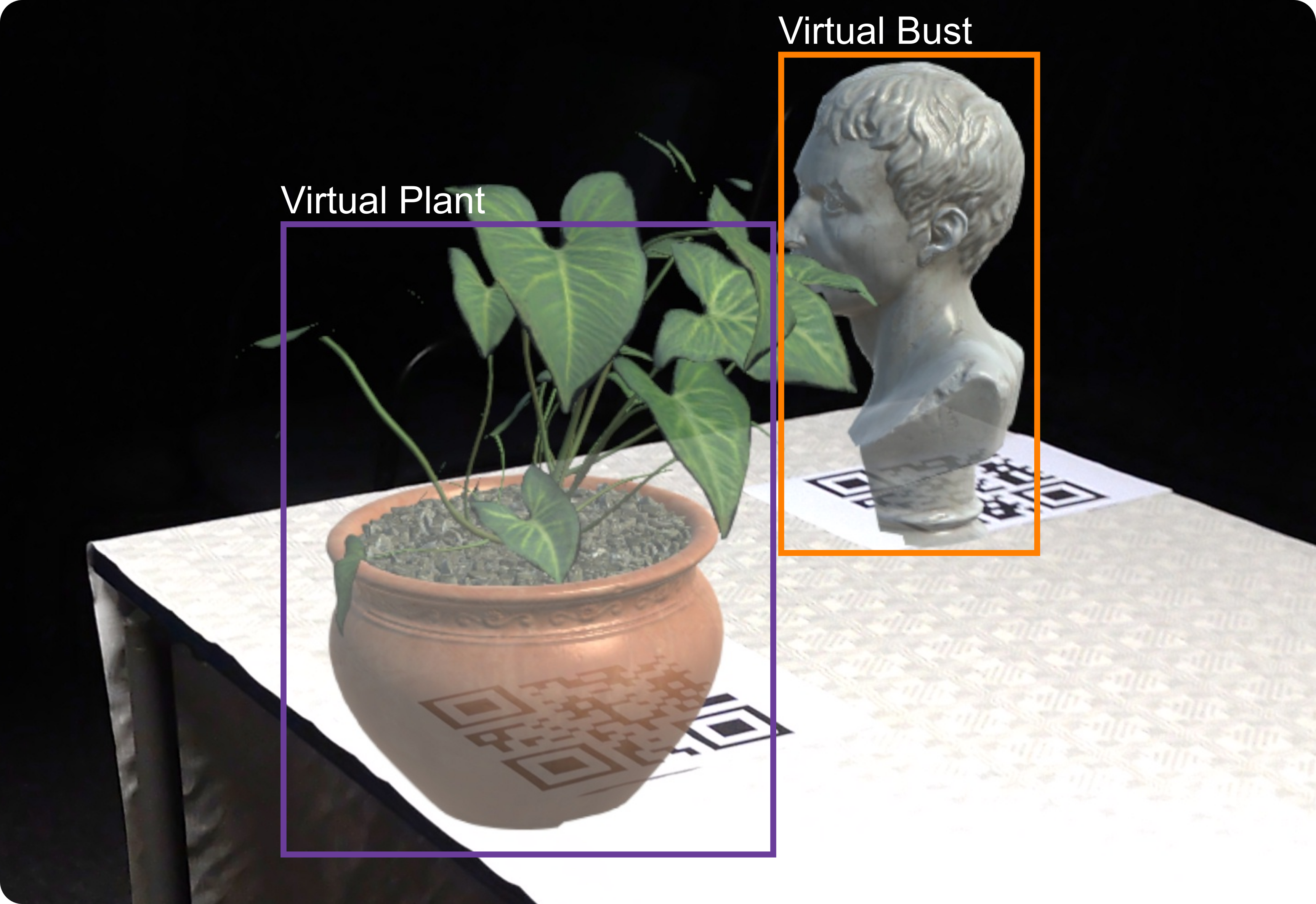}
      \caption{Virtual plant and virtual\\ bust (VP \& VB)}
      \label{fig:virtual}
    \end{subfigure}
       \begin{subfigure}[b]{0.24\textwidth}
      \includegraphics[width=\textwidth]{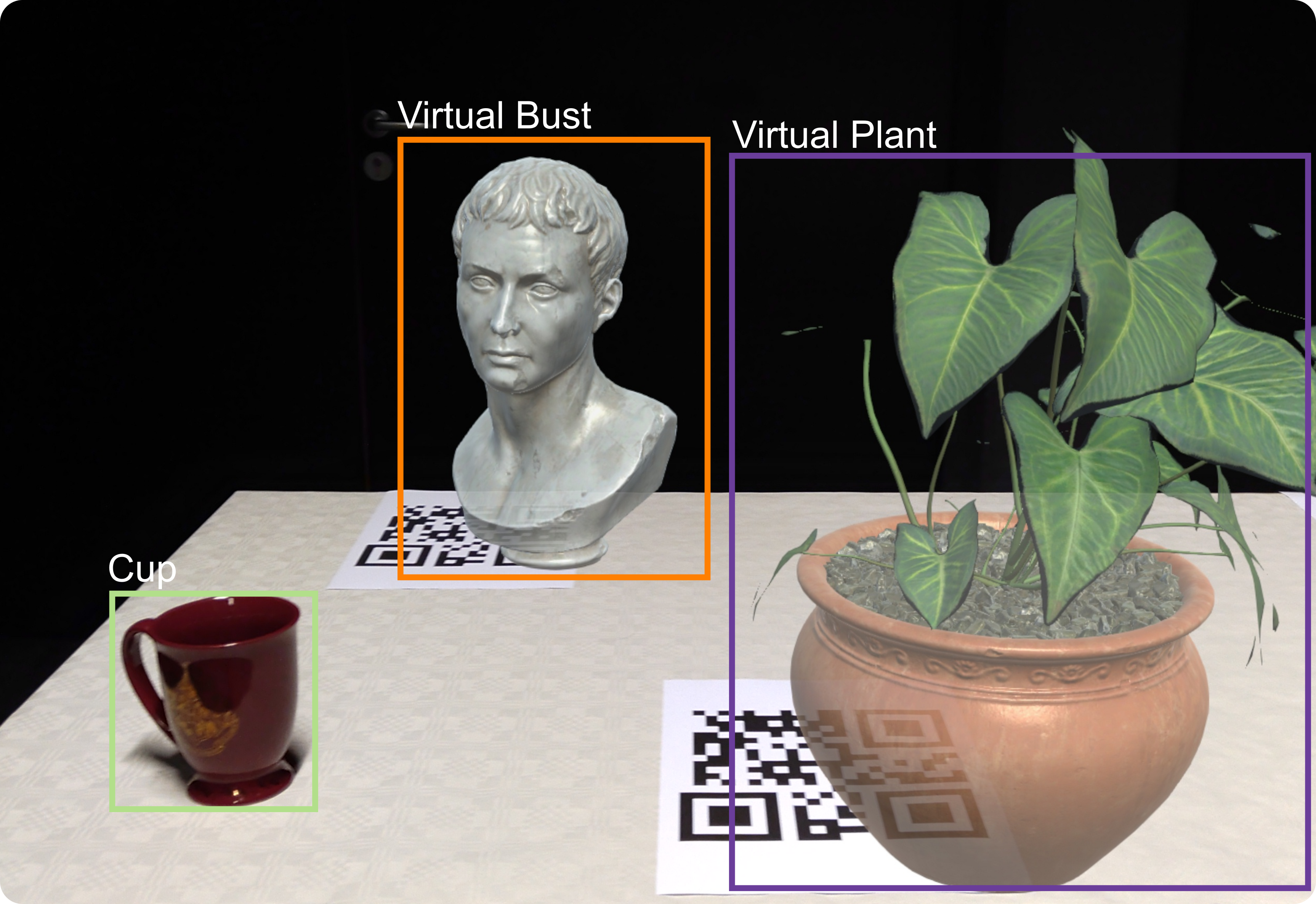}
      \caption{Virtual plant, cup and virtual bust (VP, C \& VB)}\label{fig:virtualandcup}
    \end{subfigure}
    \vspace*{-3ex}
    \Description{
    Four different images (a), (b), (c), and (d) are depicted. In the first picture, in the center, two bottles are placed diagonally, one behind the other. Both bottles are framed with a 2D blue rectangle. There is also a label on the top-left of the frame saying "bottle". In (b), There are three objects placed again diagonally, one behind the other. The objects are more on the left side of the image. Whereby the object in the front is a virtual potted plant, and behind this plant, there is a physical bottle, and behind that bottle, there is a virtual marble bust. All of these objects are also surrounded by a 2D frame, where the frame of the plant is purple, the bottle has a blue frame again, and the bust is framed in orange. (c) is similar to (b). The difference is that there is no bottle. (d) shows three objects on the table from a different perspective where the virtual plant is on the right side of the image, on the left side in the front, there is a physical cup, and at the back, there is a virtual marble bust. 
    }
    \caption{ 
    Uncertainties in stimuli: (a) The same object (bottle) at different depths. (b) A physical object (bottle) between two virtual objects (virtual plant \& virtual bust). (c) Virtual objects (virtual plant \& virtual bust) at different depths. (d) A physical object (cup) at the same depth as a virtual object (virtual plant) and another virtual object (virtual bust) at a different depth. \scriptsize(Virtual bust from guptaarnish in (a--d): \textcopyright~ Sketchfab Standard License: \url{https://sketchfab.com/3d-models/marble-bust-01-4k-c9b1839068094e40bc941b047ca19a85}).
    }\label{fig:experiment}
  \end{figure*}

\autoref{fig:algo1} illustrates an example of a \textit{standard scarf plot}, where the gaze ray intersects the cup AOI (green), but it could also happen that the participant is looking at the bottle, and considering \textit{uncertainty in order} might help identify potential ambiguities.
To represent the order of depth, we arrange the corresponding sub-segments according to their distance from the viewer from bottom to top. We refer to this extended visualization approach as a \textit{depth scarf plot} (\autoref{fig:algo2}). 
The cup AOI (green) and the bottle AOI (blue) are displayed in a single segment, with the blue sub-segment on top. To visualize the \textit{uncertainty in position}, we use the \textit{nearest neighbor (NN) scarf plot} (\autoref{fig:algo3}).  
The plot considers all AOIs within a threshold distance of the gaze for a segment.
We first identify the nearest neighbor to the gaze ray by finding the point on the ray that has a minimal distance to the center of each object. 
We then compute the Euclidean distances ($d_i$)~\cite{danielsson1980euclidean} in 3D space between this closest point on the gaze ray ($x_g$) and the center of each AOI ($x_{c_i}$). If the distance is smaller than the pre-defined threshold, we add this distance to the NN group ($G_{NN}$).
The inverse distances ($w_i$) lead to weights $p_i$.~\cite{Shepard1968}:
\begin{align}
w_i &= \begin{cases} 
      0 & d_i = 0 \\
      \frac{1}{d_i(x_g, x_{c_i})} & d_i \neq 0 
   \end{cases} \quad , \quad 
   p_i = \frac{w_i}{\sum_{j} w_j}
   \label{eq:eq1}
\end{align}

The height of each AOI in a sub-segment is based on $p_i$, and depth order is preserved in this visualization through the stacking of the sub-segments. In contrast to the \textit{standard scarf plot} and the \textit{depth scarf plot}, this plot provides more gaze-to-AOI mappings, as it also takes the neighborhood into account.

While \textit{uncertainty in order} and \textit{uncertainty in position} are associated with errors in gaze mappings, \textit{uncertainty of classification} arises from incorrect labeling due to uncertainty within the object-detection algorithm (\autoref{fig:algo4}).  
To visualize the confidence level of FP AOIs over specific time periods, we introduce a bar chart linked to segments of the scarf plot. 

\begin{table*}[htbp]
    \centering  
\caption{
Description of scarf plots shown in \autoref{fig:scarfplots}. 
\faPlusCircle\ and \faMinusCircle\ indicate the advantages and disadvantages of the visualization approaches in the respective scenes, while \faCircle\ highlights a general observation within the plots. For a full comparison of the scarf plots from all participants in each scene, we refer to the supplemental material.}
\label{tab:scarfTable}
\resizebox{\textwidth}{!}{%
    \begin{tabular}{p{1.5cm}p{4cm}p{4cm}p{4cm}p{4cm}} \hline 
    \toprule
         \textbf{Type} &  \textbf{2 bottles (BB)}& \textbf{Virtual plant, bottle, and\newline virtual bust (VP, B \& VB)}&\textbf{Virtual plant and virtual bust (VP \& VB)}& \textbf{Virtual plant, cup, and\newline virtual bust (VP, C \& VB)}\\ 
    \midrule
         \textbf{Standard scarf plot}& 
                            \faMinusCircle\ Only blue segments visible, due to identical labels (bottle) \newline  
                            \faMinusCircle\ Gaze mapping on stimulus needs to be accessed, to know to which bottle gaze is mapped 
                            & \faPlusCircle\ Task-defined visiting order is visible in the gaze pattern\newline
                            (VP $\rightarrow$ B $\rightarrow$ VB $\rightarrow$ B $\rightarrow$ VP) \newline 
                            & \faMinusCircle\ FP \textit{Book} AOI occludes \textit{VB}, leading to no gaze ray hit on \textit{VB}
                            & \faPlusCircle\ Task-defined visiting order is visible in the gaze pattern\newline
                            (VP $\rightarrow$ C $\rightarrow$ VB $\rightarrow$ C $\rightarrow$ VP)
                            \\ 
   \midrule 
         \textbf{Depth scarf plot}&
                            \faPlusCircle\ Multiple segments are split into two, revealing bottles in the foreground and background 
                            & 
                            \faPlusCircle\ Split segments show gaze ray intersection between \textit{VP} \& \textit{B}\newline and \textit{B} \& \textit{VB}
                            & \faPlusCircle\ Split segments show gaze ray intersection between FP Book AOI \& \textit{VB}
                            & 
                            \faCircle\ Similar to \textit{standard scarf plot}\newline
                            \faCircle\ Very few instances of uncertainty of order
                            \\
    \midrule 
         \textbf{NN\newline scarf plot}&
                            \faPlusCircle\ Focus shifts between\newline two bottles are visible\newline 
                            (Front bottle (bottom sub-segment larger) $\rightarrow$ bottle in the background (top sub-segment larger) $\rightarrow$ between two bottles (sub-segments have equal size))
                            & \faPlusCircle\ Focus shifts between \textit{VP} \& \textit{B} and  between \textit{B} \& \textit{VB} are visible\newline
                            & \faMinusCircle\ Multiple segments with FP \textit{Book} AOI are taking larger area within a segment 
                            & 
                            \faCircle\ Similar to \textit{standard scarf plot}\newline
                            \faCircle\ Very few instances of uncertainty of position and order
                            \\
    \bottomrule
    \end{tabular}
    }
\end{table*}
\section{Results}
We set up different scenes to demonstrate the usefulness of our uncertainty-aware scarf plots.~\autoref{fig:experiment} depicts the different scene configurations.
Each scene comprises multiple sources of uncertainty to showcase the suggested techniques in comparison with standard scarf plots as a baseline. We designed a composition of multiple virtual and physical objects that were inspected in specific static orders and from different static angles to evoke specific patterns of uncertainty. 
We used the \textit{Microsoft HoloLens~2} with the ARETT\footnote{\url{https://github.com/AR-Eye-Tracking-Toolkit/ARETT}} package by \citet{Kapp2021ARETT} and \textit{hl2\_detection}\footnote{\url{https://github.com/kolaszko/hl2_detection}} by \citet{lysakowski2023YOLO} to facilitate gaze recording and real-time object detection. We modified both projects to collect the necessary data (see supplemental material). For physical objects, we considered gaze ray intersections with their bounding boxes, whereas, for virtual objects, we accounted for direct intersections with the objects themselves.
Gaze data was recorded from four participants P1--P4 (2 male, 2 female) to compare different gaze patterns using our technique. 
\begin{figure*}[!htbp]
    \centering
    \includegraphics[width=1.0\textwidth]{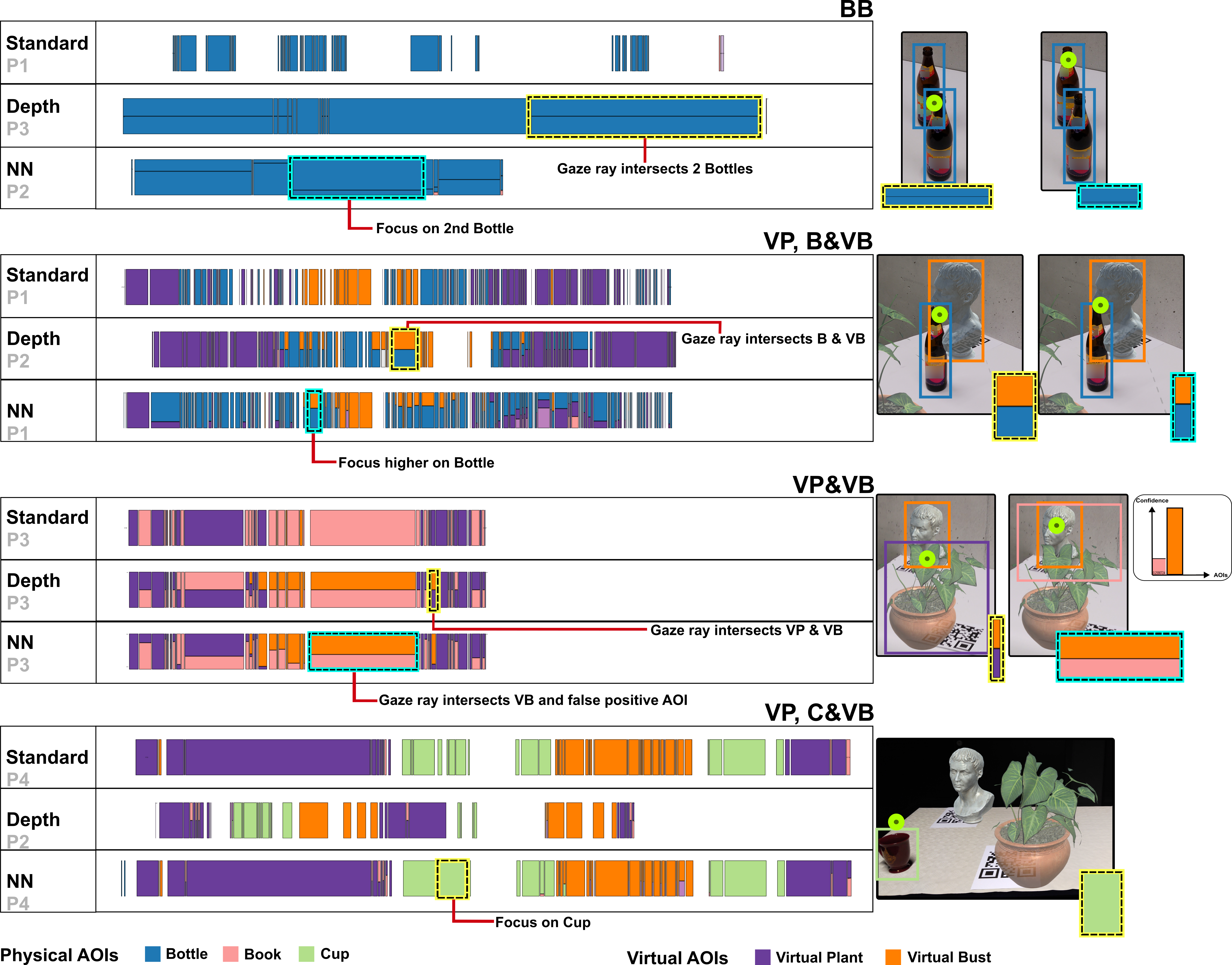}
    \Description{On the left side, there are different scarf plots for each scene, while on the right side, some stimuli from the scene are depicted with a reference segment of the scarf plots.}
    \caption{Each scene presents a scarf plot of each type based on recordings from a specific participant. While the \textit{standard scarf plot} contains limited information, the \textit{depth scarf plot} and \textit{NN scarf plot} provide additional details regarding \textit{uncertainty of order} and \textit{uncertainty of position} within each scene. In the BB scene, the blue segments in the \textit{standard scarf plot} do not reveal any insights into gaze patterns. However, the \textit{depth scarf plot} detects uncertainties of order within the scene, revealing the presence of two bottles, while the \textit{NN scarf plot} indicates which bottle the participant's focus is directed toward. In the \textit{VP, B \& VB} scene, gaze patterns caused by the task-defined visiting order can be observed in the \textit{standard scarf plot}. The \textit{depth scarf plot} and \textit{NN scarf plot} show how the gaze ray intersects the \textit{VB} and \textit{B} or \textit{VP} and \textit{B}. The \textit{VP \& VB} scene highlights the detection of a non-existent AOI (\textit{Book}), with the bar charts showing a low percentage of confidence for this AOI. While the \textit{standard scarf plot} shows no gaze on the \textit{VB} AOI due to the influence of the \textit{Book} AOI, the \textit{depth scarf plot} and \textit{NN scarf plot} reveal the gaze on \textit{VB}. In the \textit{VP, C \& VB} scene, we observe minimal uncertainties regarding order and position, resulting in similar-looking scarf plots. For a detailed comparison of each plot type using the same participant data, we refer to the supplemental material. \scriptsize(Virtual bust from guptaarnish: \textcopyright~ Sketchfab Standard License: \url{https://sketchfab.com/3d-models/marble-bust-01-4k-c9b1839068094e40bc941b047ca19a85})
    }
    \label{fig:scarfplots}
\end{figure*}
We compare the extensions \textit{depth} and \textit{NN} with the \textit{standard scarf plots} for the scenes (\autoref{tab:scarfTable}). 
\autoref{fig:scarfplots} shows the different scarf plots for each scene from selected participants. The white spaces within the scarf plots correspond to time spans without gaze data or no hits with an AOI. For a full comparison of the scarf plots from all participants in each scene, we refer to the supplemental material.
In general, our new extended plots raise awareness of uncertainties at one glance so that the data can be analyzed with more caution. While the \textit{depth scarf plot} provides information regarding objects intersecting the gaze ray at different depths, the \textit{NN scarf plot} considers objects in proximity. Therefore, it provides more segments than the \textit{standard scarf plot} and the \textit{depth scarf plot}.

As summarized in \autoref{tab:scarfTable}, the segments in the \textbf{BB} scene were solely blue for every scarf plot type since the object detection algorithm assigned identical labels to both bottles. Although the labels could be sequentially numbered based on their depth, this approach becomes impractical when multiple objects in the scene share the same label, particularly in dynamic environments. As opposed to the \textit{standard scarf plot}, the \textit{depth scarf plot} and \textit{NN scarf plot} reveal the presence of two bottles within the scene. Additionally, the \textit{NN scarf plot} provides insight into when the participant focused on each bottle, as indicated by the varying sizes of the corresponding sub-segments.

The classification algorithm occasionally produces incorrect classifications that are not prominent in the scarf plot. However, in the \textbf{VP \& VB} scene, we can notice a prominent rose-colored segment belonging to a \textit{Book} AOI, that does not exist in the scene, resulting in a FP classification.
The \textit{Book} AOI is detected in front of the \textit{VB}, therefore the \textit{standard scarf plot} does not show any gaze mapping onto the \textit{VB}, leading to unreliable results during analysis. The \textit{depth scarf plot} and \textit{NN scarf plot} reveal that the gaze ray intersected both the FP \textit{Book} AOI and \textit{VB}, resulting in more reliable analysis outcomes. By selecting the segments that include the FP \textit{Book} AOI, we can inspect within the linked bar chart the confidence of classification to confirm the misclassification. If a classification error is prominent, such as the FP \textit{Book} AOI in our example, the analyst may consider re-visualizing the scarf plot without the FP AOIs in the dataset.  
A version without the \textit{Book} AOI is available in the supplemental material.  

In the \textbf{VP, B \& VB} scene, the gaze patterns in the \textit{standard scarf plot} show the task-defined visiting order. While this order is still visible in the \textit{depth scarf plot} and the \textit{NN scarf plot}, we can additionally see that the \textit{B} is close to the \textit{VB} and the \textit{VP}, providing segments that are split into \textit{VP} \& \textit{B} and \textit{B} \& \textit{VB}. The \textit{NN scarf plot}, additionally, shows that the participant focused the majority of time on the \textit{B}. 

We also noticed how some scene setups were more prone to uncertainties than others. For example, the \textbf{VP \& VB} and the \textbf{VP, C~\&~VB} scenes had fewer uncertainties regarding order and position, than the other two scenes. 

\section{Discussion}
Our demonstration outlined how uncertainty-aware scarf plots can show the most common potential issues of uncertainty in eye-tracking data occurring mainly through positional offset, depth, and automatic detection of AOIs. 

\subsection{Potential Application Scenarios}
Scarf plots can also aid in improving the design of an environment, such that uncertainties of order and position of objects are mitigated. 
Since not all environments can be redesigned, some uncertainties are unavoidable. 
In such cases, uncertainty-aware scarf plots can help extract gaze patterns or make alternative gaze sequences visible. 
Additionally, these plots can be included in supplemental materials when reporting results regarding gaze patterns.

\paragraph{Occlusion}
Our technique can be beneficial in various AR applications, with multiple physical and virtual objects. The \textit{standard scarf plot} may produce unreliable results when the scene contains multiple occluding objects. This could happen in museum exhibitions in AR~\cite{Jin2024}, where the physical artworks get extended by virtual content, or during fabrication tasks extended by virtual content to aid the worker~\cite{Becher22}. In both scenarios, the \textit{standard scarf plot} will only yield the segment intersected first by the gaze ray. The \textit{depth scarf plot} makes the analyst aware of the physical object behind the virtual content, enabling a more reliable interpretation. 
The \textit{NN scarf plot} reveals whether the user was more likely looking at the physical or virtual object. 

\paragraph{Collaboration}
The extended scarf plots can also be beneficial when analyzing shared attention of collaborative tasks in AR~\cite{Pathmanathan24}. 
Depending on their perspective, one user might look through a virtual object at a physical one, while another user views the same physical object directly. A \textit{standard scarf plot} would display different segments for each user, whereas our techniques would indicate that the gaze ray of both users intersects two objects, allowing the analysts to assume that both users are looking at the same object.  

\paragraph{Visual search task}
When users perform visual search tasks, for instance, for testing attention guidance techniques~\cite{Renner17}, the scene is often crowded by objects that are close to each other and overlap, leading to uncertainties in position and order. When evaluating an attention guidance technique, however, the user will follow a specific pattern with their gaze, which should be visible in the scarf plot. The \textit{NN scarf plot} 
can show the expected gaze pattern when only the sub-segments with the largest area are considered (see supplemental material).
\paragraph{Object detection}
When analyzing user perception, for instance, in guide applications that use object detection to place virtual content on a physical object~\cite{Pawade2018} or playback certain audio based on the detected object~\cite{Kaul21}, the analyst would be able to inspect the confidence of detected labels at time spans, where the sequence of segments deviates from expectations. If such classification errors happen more often during the analysis, the analyst can consider improving the classifier. Otherwise, the corresponding segments can be manually adapted.  
\subsection{Limitations}
Scarf plots are generally known for their color processing limits, i.e., poor visual scalability \cite{Richer:2024:scalability} with increasing number of AOIs. In our case, the larger number of AOIs within a segment can further amplify this issue. Fortunately, the number of AOIs visible within a user's peripheral vision is naturally limited. By setting an appropriate threshold for the NN approach and the number of objects intersected by the gaze ray, we can filter out unnecessary AOIs. 
Furthermore, FP classification errors cannot be detected directly, particularly when scene objects could potentially correspond to the given label. Identifying such mistakes would require more in-depth analysis. 
Further limitations include sources of uncertainty that our approach does not address. A detailed description of these uncertainties is provided in the supplemental material. Besides, gaze recorded in dynamic settings could introduce additional limitations not covered in this work.

\section{Conclusion}
We presented a new approach to include the most common aspects of uncertainty in eye-tracking data in a visualization. The uncertainty-aware scarf plots do not only allow the detection of patterns in gaze data but also highlight where potential ambiguities in the mapping between gaze and AOIs exist. To demonstrate our technique, we created a scene that contains the aforementioned uncertainties and illustrated how the corresponding scarf plots visualize them. Furthermore, we discussed how our scarf plots can be applied in specific environments and scenarios, as well as the limitations of our technique.
In general, our approach allows the analysts to reason about possible misunderstandings in the interpretation of gaze patterns and get a better overview of the perceptional behavior of participants. 

Our work aims to inform researchers about uncertainty in the data of their experiments. 
We think that the proposed visualization helps identify issues in recording quality as well as with study designs and take appropriate countermeasures early on, for instance, during pilot testing. 
Additionally, such visualizations can serve as supplemental material to statistical analyses to support findings and explain potential limitations. 
While our work focuses only on visualizing uncertainties of order, position, and classification, it can be extended by incorporating representations of further uncertainties (e.g., low sampling rate, calibration drifts, synchronization issues) in a multi-coordinated view approach.

\subsection*{Ethical Considerations and Privacy}
General privacy concerns related to personal data identifiable from eye-tracking data are a potential risk for extended scarf plots that might provide more details than a standard scarf plot.  

\begin{acks}
This work is supported by the \grantsponsor{DFG}{Deutsche Forschungsgemeinschaft (DFG, German Research Foundation)}{https://www.dfg.de/} under Germany’s Excellence Strategy -- EXC 2120/1 -- \grantnum{DFG}{390831618}, Project-ID \grantnum{DFG}{449742818}, and Project-ID \grantnum{DFG}{251654672} -- TRR 161.
\end{acks}
\bibliographystyle{ACM-Reference-Format}

\begin{thebibliography}{47}


\ifx \showCODEN    \undefined \def \showCODEN     #1{\unskip}     \fi
\ifx \showDOI      \undefined \def \showDOI       #1{#1}\fi
\ifx \showISBNx    \undefined \def \showISBNx     #1{\unskip}     \fi
\ifx \showISBNxiii \undefined \def \showISBNxiii  #1{\unskip}     \fi
\ifx \showISSN     \undefined \def \showISSN      #1{\unskip}     \fi
\ifx \showLCCN     \undefined \def \showLCCN      #1{\unskip}     \fi
\ifx \shownote     \undefined \def \shownote      #1{#1}          \fi
\ifx \showarticletitle \undefined \def \showarticletitle #1{#1}   \fi
\ifx \showURL      \undefined \def \showURL       {\relax}        \fi
\providecommand\bibfield[2]{#2}
\providecommand\bibinfo[2]{#2}
\providecommand\natexlab[1]{#1}
\providecommand\showeprint[2][]{arXiv:#2}

\bibitem[Andersson et~al\mbox{.}(2010)]%
        {Andersson2010SamplingFreq}
\bibfield{author}{\bibinfo{person}{Richard Andersson}, \bibinfo{person}{Marcus Nystr{\"o}m}, {and} \bibinfo{person}{Kenneth Holmqvist}.} \bibinfo{year}{2010}\natexlab{}.
\newblock \showarticletitle{Sampling {{Frequency}} and {{Eye-Tracking Measures}}: {{How Speed Affects Durations}}, {{Latencies}}, and More}.
\newblock \bibinfo{journal}{\emph{Journal of Eye Movement Research}} \bibinfo{volume}{3}, \bibinfo{number}{3} (\bibinfo{year}{2010}), \bibinfo{pages}{1--12}.
\newblock
\urldef\tempurl%
\url{https://doi.org/10.16910/jemr.3.3.6}
\showDOI{\tempurl}


\bibitem[Andrienko et~al\mbox{.}(2012)]%
        {andrienko2012visual}
\bibfield{author}{\bibinfo{person}{Gennady Andrienko}, \bibinfo{person}{Natalia Andrienko}, \bibinfo{person}{Michael Burch}, {and} \bibinfo{person}{Daniel Weiskopf}.} \bibinfo{year}{2012}\natexlab{}.
\newblock \showarticletitle{Visual {{Analytics Methodology}} for {{Eye Movement Studies}}}.
\newblock \bibinfo{journal}{\emph{IEEE Transactions on Visualization and Computer Graphics}} \bibinfo{volume}{18}, \bibinfo{number}{12} (\bibinfo{year}{2012}), \bibinfo{pages}{2889--2898}.
\newblock
\urldef\tempurl%
\url{https://doi.org/10.1109/TVCG.2012.276}
\showDOI{\tempurl}


\bibitem[Andrienko et~al\mbox{.}(2022)]%
        {andrienko2022seeking}
\bibfield{author}{\bibinfo{person}{Natalia Andrienko}, \bibinfo{person}{Gennady. Andrienko}, \bibinfo{person}{Simming Chen}, {and} \bibinfo{person}{Brian Fisher}.} \bibinfo{year}{2022}\natexlab{}.
\newblock \showarticletitle{Seeking Patterns of Visual Pattern Discovery for Knowledge Building}.
\newblock \bibinfo{journal}{\emph{Computer Graphics Forum}} \bibinfo{volume}{41}, \bibinfo{number}{6} (\bibinfo{year}{2022}), \bibinfo{pages}{124--148}.
\newblock
\urldef\tempurl%
\url{https://doi.org/10.1111/cgf.14515}
\showDOI{\tempurl}


\bibitem[Barz et~al\mbox{.}(2023)]%
        {barz2023interactive}
\bibfield{author}{\bibinfo{person}{Michael Barz}, \bibinfo{person}{Omair~Shahzad Bhatti}, \bibinfo{person}{Hasan Md~Tusfiqur Alam}, \bibinfo{person}{Duy Minh~Ho Nguyen}, {and} \bibinfo{person}{Daniel Sonntag}.} \bibinfo{year}{2023}\natexlab{}.
\newblock \showarticletitle{Interactive {{Fixation-to-AOI Mapping}} for {{Mobile Eye Tracking Data Based}} on {{Few-Shot Image Classification}}}. In \bibinfo{booktitle}{\emph{{{International Conference}} on {{Intelligent User Interfaces}}}}. \bibinfo{publisher}{ACM}, \bibinfo{pages}{175--178}.
\newblock
\showISBNx{979-8-4007-0107-8}
\urldef\tempurl%
\url{https://doi.org/10.1145/3581754.3584179}
\showDOI{\tempurl}


\bibitem[Becher et~al\mbox{.}(2022)]%
        {Becher22}
\bibfield{author}{\bibinfo{person}{Michael Becher}, \bibinfo{person}{Dominik Herr}, \bibinfo{person}{Christoph Muller}, \bibinfo{person}{Kuno Kurzhals}, \bibinfo{person}{Guido Reina}, \bibinfo{person}{Lena Wagner}, \bibinfo{person}{Thomas Ertl}, {and} \bibinfo{person}{Daniel Weiskopf}.} \bibinfo{year}{2022}\natexlab{}.
\newblock \showarticletitle{Situated {{Visual Analysis}} and {{Live Monitoring}} for {{Manufacturing}}}.
\newblock \bibinfo{journal}{\emph{IEEE Computer Graphics and Applications}} \bibinfo{volume}{42}, \bibinfo{number}{02} (\bibinfo{year}{2022}), \bibinfo{pages}{33--44}.
\newblock
\urldef\tempurl%
\url{https://doi.org/10.1109/MCG.2022.3157961}
\showDOI{\tempurl}


\bibitem[Blascheck et~al\mbox{.}(2017)]%
        {blascheck2017taxonomy}
\bibfield{author}{\bibinfo{person}{Tanja Blascheck}, \bibinfo{person}{Kuno Kurzhals}, \bibinfo{person}{Michael Raschke}, \bibinfo{person}{Michael Burch}, \bibinfo{person}{Daniel Weiskopf}, {and} \bibinfo{person}{T. Ertl}.} \bibinfo{year}{2017}\natexlab{}.
\newblock \showarticletitle{Visualization of Eye Tracking Data: A Taxonomy and Survey}.
\newblock \bibinfo{journal}{\emph{Computer Graphics Forum}} \bibinfo{volume}{36}, \bibinfo{number}{8} (\bibinfo{year}{2017}), \bibinfo{pages}{260--284}.
\newblock
\urldef\tempurl%
\url{https://doi.org/10.1111/cgf.13079}
\showDOI{\tempurl}


\bibitem[Bonneau et~al\mbox{.}(2014)]%
        {Bonneau:2014:OSU}
\bibfield{author}{\bibinfo{person}{Georges-Pierre Bonneau}, \bibinfo{person}{Hans-Christian Hege}, \bibinfo{person}{Chris~R. Johnson}, \bibinfo{person}{Manuel~M. Oliveira}, \bibinfo{person}{Kristin Potter}, \bibinfo{person}{Penny Rheingans}, {and} \bibinfo{person}{Thomas Schultz}.} \bibinfo{year}{2014}\natexlab{}.
\newblock \showarticletitle{Overview and {{State-of-the-Art}} of {{Uncertainty Visualization}}}.
\newblock In \bibinfo{booktitle}{\emph{Scientific {{Visualization}}}}, \bibfield{editor}{\bibinfo{person}{Charles~D. Hansen}, \bibinfo{person}{Min Chen}, \bibinfo{person}{Christopher~R. Johnson}, \bibinfo{person}{Arie~E. Kaufman}, {and} \bibinfo{person}{Hans Hagen}} (Eds.). \bibinfo{publisher}{Springer}, \bibinfo{pages}{3--27}.
\newblock
\urldef\tempurl%
\url{https://doi.org/10.1007/978-1-4471-6497-5_1}
\showDOI{\tempurl}


\bibitem[Danielsson(1980)]%
        {danielsson1980euclidean}
\bibfield{author}{\bibinfo{person}{Per-Erik Danielsson}.} \bibinfo{year}{1980}\natexlab{}.
\newblock \showarticletitle{Euclidean {{Distance Mapping}}}.
\newblock \bibinfo{journal}{\emph{Computer Graphics and Image Processing}} \bibinfo{volume}{14}, \bibinfo{number}{3} (\bibinfo{year}{1980}), \bibinfo{pages}{227--248}.
\newblock
\urldef\tempurl%
\url{https://doi.org/10.1016/0146-664X(80)90054-4}
\showDOI{\tempurl}


\bibitem[De~Beugher et~al\mbox{.}(2012)]%
        {de2012automatic}
\bibfield{author}{\bibinfo{person}{Stijn De~Beugher}, \bibinfo{person}{Younes Ichiche}, \bibinfo{person}{Geert Br{\^o}ne}, {and} \bibinfo{person}{Toon Goedem{\'e}}.} \bibinfo{year}{2012}\natexlab{}.
\newblock \showarticletitle{Automatic {{Analysis}} of {{Eye-Tracking Data}} Using {{Object Detection Algorithms}}}. In \bibinfo{booktitle}{\emph{{{Conference}} on {{Ubiquitous Computing}}}}. \bibinfo{publisher}{ACM}, \bibinfo{pages}{677--680}.
\newblock
\showISBNx{978-1-4503-1224-0}
\urldef\tempurl%
\url{https://doi.org/10.1145/2370216.2370363}
\showDOI{\tempurl}


\bibitem[Deane et~al\mbox{.}(2023)]%
        {deane2023deep}
\bibfield{author}{\bibinfo{person}{Oliver Deane}, \bibinfo{person}{Eszter Toth}, {and} \bibinfo{person}{Sang-Hoon Yeo}.} \bibinfo{year}{2023}\natexlab{}.
\newblock \showarticletitle{Deep-{{SAGA}}: {{A Deep-Learning-Based System}} for {{Automatic Gaze Annotation}} from {{Eye-Tracking Data}}}.
\newblock \bibinfo{journal}{\emph{Behavior Research Methods}} \bibinfo{volume}{55}, \bibinfo{number}{3} (\bibinfo{year}{2023}), \bibinfo{pages}{1372--1391}.
\newblock
\urldef\tempurl%
\url{https://doi.org/10.3758/s13428-022-01833-4}
\showDOI{\tempurl}


\bibitem[Dzsotjan et~al\mbox{.}(2021)]%
        {dzsotjan2021predictive}
\bibfield{author}{\bibinfo{person}{David Dzsotjan}, \bibinfo{person}{Kim {Ludwig-Petsch}}, \bibinfo{person}{Sergey Mukhametov}, \bibinfo{person}{Shoya Ishimaru}, \bibinfo{person}{Stefan Kuechemann}, {and} \bibinfo{person}{Jochen Kuhn}.} \bibinfo{year}{2021}\natexlab{}.
\newblock \showarticletitle{The {{Predictive Power}} of {{Eye-Tracking Data}} in an {{Interactive AR Learning Environment}}}. In \bibinfo{booktitle}{\emph{{{International Joint Conference}} on {{Pervasive}} and {{Ubiquitous Computing}} and {{International Symposium}} on {{Wearable Computers}}}}. \bibinfo{publisher}{ACM}, \bibinfo{pages}{467--471}.
\newblock
\showISBNx{978-1-4503-8461-2}
\urldef\tempurl%
\url{https://doi.org/10.1145/3460418.3479358}
\showDOI{\tempurl}


\bibitem[Falzone et~al\mbox{.}(2023)]%
        {falzone2023using}
\bibfield{author}{\bibinfo{person}{Kevin Falzone}, \bibinfo{person}{Sophie Lemonnier}, \bibinfo{person}{Thibaut Gr{\'e}bert}, {and} \bibinfo{person}{Christian Bastien}.} \bibinfo{year}{2023}\natexlab{}.
\newblock \showarticletitle{Using {{Scarf Plots}} to {{Visualize Moment-to-Moment Visual Search Behavior}} on {{Websites}}}. In \bibinfo{booktitle}{\emph{International {{Francophone Conference}} on {{Human-Computer Interaction}}}}. \bibinfo{publisher}{ACM}, \bibinfo{pages}{1--8}.
\newblock
\showISBNx{979-8-4007-0066-8}
\urldef\tempurl%
\url{https://doi.org/10.1145/3577590.3589604}
\showDOI{\tempurl}


\bibitem[Griethe and Schumann(2006)]%
        {Griethe:2006:VUD}
\bibfield{author}{\bibinfo{person}{Henning Griethe} {and} \bibinfo{person}{Heidrun Schumann}.} \bibinfo{year}{2006}\natexlab{}.
\newblock \showarticletitle{The {{Visualization}} of {{Uncertain Data}}: {{Methods}} and {{Problems}}}. In \bibinfo{booktitle}{\emph{Simulation and {{Visualization}}}}. \bibinfo{publisher}{SCS Publishing House}, \bibinfo{pages}{143--156}.
\newblock


\bibitem[Holmqvist et~al\mbox{.}(2011)]%
        {holmqvist2011eye}
\bibfield{author}{\bibinfo{person}{Kenneth Holmqvist}, \bibinfo{person}{Marcus Nystr{\"o}m}, \bibinfo{person}{Richard Andersson}, \bibinfo{person}{Richard Dewhurst}, \bibinfo{person}{Halszka Jarodzka}, {and} \bibinfo{person}{Joost {Van de Weijer}}.} \bibinfo{year}{2011}\natexlab{}.
\newblock \bibinfo{booktitle}{\emph{Eye {{Tracking}}: A {{Comprehensive Guide}} to {{Methods}} and {{Measures}}}}.
\newblock \bibinfo{publisher}{Oxford University Press}, \bibinfo{address}{Oxford}.
\newblock
\showISBNx{978-0-19-969708-3}


\bibitem[Holmqvist et~al\mbox{.}(2012)]%
        {holmqvist2012eye}
\bibfield{author}{\bibinfo{person}{Kenneth Holmqvist}, \bibinfo{person}{Marcus Nystr{\"o}m}, {and} \bibinfo{person}{Fiona Mulvey}.} \bibinfo{year}{2012}\natexlab{}.
\newblock \showarticletitle{Eye {{Tracker Data Quality}}: {{What}} It Is and {{How}} to {{Measure}} It}. In \bibinfo{booktitle}{\emph{{{ACM Symposium}} on {{Eye Tracking Research}} and {{Applications}}}}. \bibinfo{publisher}{ACM}, \bibinfo{pages}{45--52}.
\newblock
\showISBNx{978-1-4503-1221-9}
\urldef\tempurl%
\url{https://doi.org/10.1145/2168556.2168563}
\showDOI{\tempurl}


\bibitem[Jin et~al\mbox{.}(2024)]%
        {Jin2024}
\bibfield{author}{\bibinfo{person}{Yunshui Jin}, \bibinfo{person}{Minhua Ma}, {and} \bibinfo{person}{Yun Liu}.} \bibinfo{year}{2024}\natexlab{}.
\newblock \showarticletitle{Comparative {{Study}} of {{HMD-based Virtual}} and {{Augmented Realities}} for {{Immersive Museums}}: {{User Acceptance}}, {{Medium}}, and {{Learning}}}.
\newblock \bibinfo{journal}{\emph{ACM Journal on Computing and Cultural Heritage}} \bibinfo{volume}{17}, \bibinfo{number}{1} (\bibinfo{year}{2024}), \bibinfo{pages}{1--17}.
\newblock
\urldef\tempurl%
\url{https://doi.org/10.1145/3627164}
\showDOI{\tempurl}


\bibitem[Jocher et~al\mbox{.}(2023)]%
        {yolov8_ultralytics}
\bibfield{author}{\bibinfo{person}{Glenn Jocher}, \bibinfo{person}{Jing Qiu}, {and} \bibinfo{person}{Ayush Chaurasia}.} \bibinfo{year}{2023}\natexlab{}.
\newblock \bibinfo{booktitle}{\emph{{Ultralytics YOLO}}}.
\newblock
\urldef\tempurl%
\url{https://github.com/ultralytics/ultralytics}
\showURL{%
\tempurl}


\bibitem[Kapp et~al\mbox{.}(2021)]%
        {Kapp2021ARETT}
\bibfield{author}{\bibinfo{person}{Sebastian Kapp}, \bibinfo{person}{Michael Barz}, \bibinfo{person}{Sergey Mukhametov}, \bibinfo{person}{Daniel Sonntag}, {and} \bibinfo{person}{Jochen Kuhn}.} \bibinfo{year}{2021}\natexlab{}.
\newblock \showarticletitle{{{ARETT}}: {{Augmented Reality Eye Tracking Toolkit}} for {{Head Mounted Displays}}}.
\newblock \bibinfo{journal}{\emph{Sensors}} \bibinfo{volume}{21}, \bibinfo{number}{6} (\bibinfo{year}{2021}), \bibinfo{pages}{2234}.
\newblock
\showISSN{1424-8220}
\urldef\tempurl%
\url{https://doi.org/10.3390/s21062234}
\showDOI{\tempurl}


\bibitem[Kaul et~al\mbox{.}(2021)]%
        {Kaul21}
\bibfield{author}{\bibinfo{person}{Oliver~Beren Kaul}, \bibinfo{person}{Kersten Behrens}, {and} \bibinfo{person}{Michael Rohs}.} \bibinfo{year}{2021}\natexlab{}.
\newblock \showarticletitle{Mobile {{Recognition}} and {{Tracking}} of {{Objects}} in the {{Environment}} through {{Augmented Reality}} and {{3D Audio Cues}} for {{People}} with {{Visual Impairments}}}. In \bibinfo{booktitle}{\emph{Extended {{Abstracts}} of the {{Conference}} on {{Human Factors}} in {{Computing Systems}}}}. \bibinfo{publisher}{ACM}, \bibinfo{pages}{1--7}.
\newblock
\showISBNx{978-1-4503-8095-9}
\urldef\tempurl%
\url{https://doi.org/10.1145/3411763.3451611}
\showDOI{\tempurl}


\bibitem[Koch et~al\mbox{.}(2024)]%
        {Koch2024ActiveGaze}
\bibfield{author}{\bibinfo{person}{Maurice Koch}, \bibinfo{person}{Nan Cao}, \bibinfo{person}{Daniel Weiskopf}, {and} \bibinfo{person}{Kuno Kurzhals}.} \bibinfo{year}{2024}\natexlab{}.
\newblock \showarticletitle{Active {{Gaze Labeling}}: {{Visualization}} for {{Trust Building}}}.
\newblock \bibinfo{journal}{\emph{IEEE Transactions on Visualization and Computer Graphics}} (\bibinfo{year}{2024}), \bibinfo{pages}{1--15}.
\newblock
\urldef\tempurl%
\url{https://doi.org/10.1109/TVCG.2024.3392476}
\showDOI{\tempurl}


\bibitem[Kop{\'a}csi et~al\mbox{.}(2023)]%
        {kopacsi2023imeta}
\bibfield{author}{\bibinfo{person}{L{\'a}szl{\'o} Kop{\'a}csi}, \bibinfo{person}{Michael Barz}, \bibinfo{person}{Omair~Shahzad Bhatti}, {and} \bibinfo{person}{Daniel Sonntag}.} \bibinfo{year}{2023}\natexlab{}.
\newblock \showarticletitle{{{IMETA}}: {{An Interactive Mobile Eye Tracking Annotation Method}} for {{Semi-Automatic Fixation-to-AOI}} Mapping}. In \bibinfo{booktitle}{\emph{International {{Conference}} on {{Intelligent User Interfaces}}}}. \bibinfo{publisher}{ACM}, \bibinfo{pages}{33--36}.
\newblock
\showISBNx{979-8-4007-0107-8}
\urldef\tempurl%
\url{https://doi.org/10.1145/3581754.3584125}
\showDOI{\tempurl}


\bibitem[Kosslyn(1978)]%
        {KosslynVisualAngle}
\bibfield{author}{\bibinfo{person}{Stephen~Michael Kosslyn}.} \bibinfo{year}{1978}\natexlab{}.
\newblock \showarticletitle{Measuring the {{Visual Angle}} of the {{Mind}}'s {{Eye}}}.
\newblock \bibinfo{journal}{\emph{Cognitive Psychology}} \bibinfo{volume}{10}, \bibinfo{number}{3} (\bibinfo{year}{1978}), \bibinfo{pages}{356--389}.
\newblock
\urldef\tempurl%
\url{https://doi.org/10.1016/0010-0285(78)90004-X}
\showDOI{\tempurl}


\bibitem[Kurzhals et~al\mbox{.}(2022)]%
        {kurzhals2022Situated}
\bibfield{author}{\bibinfo{person}{Kuno Kurzhals}, \bibinfo{person}{Michael Becher}, \bibinfo{person}{Nelusa Pathmanathan}, {and} \bibinfo{person}{Guido Reina}.} \bibinfo{year}{2022}\natexlab{}.
\newblock \showarticletitle{Evaluating {{Situated Visualization}} in {{AR}} with {{Eye Tracking}}}. In \bibinfo{booktitle}{\emph{{{Evaluation}} and {{Beyond}} - {{Methodological Approaches}} for {{Visualization}}}}. \bibinfo{publisher}{IEEE}, \bibinfo{pages}{77--84}.
\newblock
\urldef\tempurl%
\url{https://doi.org/10.1109/BELIV57783.2022.00013}
\showDOI{\tempurl}


\bibitem[Kurzhals et~al\mbox{.}(2014)]%
        {kurzhals2014iseecube}
\bibfield{author}{\bibinfo{person}{Kuno Kurzhals}, \bibinfo{person}{Florian Heimerl}, {and} \bibinfo{person}{Daniel Weiskopf}.} \bibinfo{year}{2014}\natexlab{}.
\newblock \showarticletitle{{{ISeeCube}}: {{Visual Analysis}} of {{Gaze Data}} for {{Video}}}. In \bibinfo{booktitle}{\emph{{{Symposium}} on {{Eye Tracking Research}} and {{Applications}}}}. \bibinfo{publisher}{ACM}, \bibinfo{pages}{43--50}.
\newblock
\showISBNx{978-1-4503-2751-0}
\urldef\tempurl%
\url{https://doi.org/10.1145/2578153.2578158}
\showDOI{\tempurl}


\bibitem[Kurzhals et~al\mbox{.}(2017)]%
        {Kurzhals2017Clustering}
\bibfield{author}{\bibinfo{person}{Kuno Kurzhals}, \bibinfo{person}{Marcel Hlawatsch}, \bibinfo{person}{Christof Seeger}, {and} \bibinfo{person}{Daniel Weiskopf}.} \bibinfo{year}{2017}\natexlab{}.
\newblock \showarticletitle{Visual {{Analytics}} for {{Mobile Eye Tracking}}}.
\newblock \bibinfo{journal}{\emph{IEEE Transactions on Visualization and Computer Graphics}} \bibinfo{volume}{23}, \bibinfo{number}{1} (\bibinfo{year}{2017}), \bibinfo{pages}{301--310}.
\newblock
\urldef\tempurl%
\url{https://doi.org/10.1109/TVCG.2016.2598695}
\showDOI{\tempurl}


\bibitem[{\L}ysakowski et~al\mbox{.}(2023)]%
        {lysakowski2023YOLO}
\bibfield{author}{\bibinfo{person}{Miko{\l}aj {\L}ysakowski}, \bibinfo{person}{Kamil {\.Z}ywanowski}, \bibinfo{person}{Adam Banaszczyk}, \bibinfo{person}{Micha{\l}~R. Nowicki}, \bibinfo{person}{Piotr Skrzypczy{\'n}ski}, {and} \bibinfo{person}{S{\l}awomir~K. Tadeja}.} \bibinfo{year}{2023}\natexlab{}.
\newblock \showarticletitle{Real-{{Time Onboard Object Detection}} for {{Augmented Reality}}: {{Enhancing Head-Mounted Display}} with {{YOLOv8}}}. In \bibinfo{booktitle}{\emph{{{International Conference}} on {{Edge Computing}} and {{Communications}}}}. \bibinfo{publisher}{IEEE}, \bibinfo{pages}{364--371}.
\newblock
\urldef\tempurl%
\url{https://doi.org/10.1109/EDGE60047.2023.00059}
\showDOI{\tempurl}


\bibitem[{\"O}ney et~al\mbox{.}(2023)]%
        {oney2023_AR}
\bibfield{author}{\bibinfo{person}{Seyda {\"O}ney}, \bibinfo{person}{Nelusa Pathmanathan}, \bibinfo{person}{Michael Becher}, \bibinfo{person}{Michael Sedlmair}, \bibinfo{person}{Daniel Weiskopf}, {and} \bibinfo{person}{Kuno Kurzhals}.} \bibinfo{year}{2023}\natexlab{}.
\newblock \showarticletitle{Visual {{Gaze Labeling}} for {{Augmented Reality Studies}}}. In \bibinfo{booktitle}{\emph{Computer {{Graphics Forum}}}}, Vol.~\bibinfo{volume}{42}. \bibinfo{publisher}{Wiley Online Library}, \bibinfo{pages}{373--384}.
\newblock
\urldef\tempurl%
\url{https://doi.org/10.1111/cgf.14837}
\showDOI{\tempurl}


\bibitem[Padilla et~al\mbox{.}(2021)]%
        {Padilla:2020:UV}
\bibfield{author}{\bibinfo{person}{Lace Padilla}, \bibinfo{person}{Matthew Kay}, {and} \bibinfo{person}{Jessica Hullman}.} \bibinfo{year}{2021}\natexlab{}.
\newblock \bibinfo{booktitle}{\emph{Uncertainty {{Visualization}}}}.
\newblock \bibinfo{publisher}{John Wiley \& Sons, Ltd}, \bibinfo{pages}{1--18}.
\newblock
\showISBNx{9781118445112}
\urldef\tempurl%
\url{https://doi.org/10.1002/9781118445112.stat08296}
\showDOI{\tempurl}


\bibitem[Panetta et~al\mbox{.}(2019)]%
        {panetta2019software}
\bibfield{author}{\bibinfo{person}{Karen Panetta}, \bibinfo{person}{Qianwen Wan}, \bibinfo{person}{Aleksandra Kaszowska}, \bibinfo{person}{Holly~A Taylor}, {and} \bibinfo{person}{Sos Agaian}.} \bibinfo{year}{2019}\natexlab{}.
\newblock \showarticletitle{Software {{Architecture}} for {{Automating Cognitive Science Eye-Tracking Data Analysis}} and {{Object Annotation}}}.
\newblock \bibinfo{journal}{\emph{IEEE Transactions on Human-Machine Systems}} \bibinfo{volume}{49}, \bibinfo{number}{3} (\bibinfo{year}{2019}), \bibinfo{pages}{268--277}.
\newblock
\urldef\tempurl%
\url{https://doi.org/10.1109/THMS.2019.2892919}
\showDOI{\tempurl}


\bibitem[Pathmanathan et~al\mbox{.}(2024)]%
        {Pathmanathan24}
\bibfield{author}{\bibinfo{person}{Nelusa Pathmanathan}, \bibinfo{person}{Tobias Rau}, \bibinfo{person}{Xiliu Yang}, \bibinfo{person}{Aim{\'e}e~Sousa Calepso}, \bibinfo{person}{Felix Amtsberg}, \bibinfo{person}{Achim Menges}, \bibinfo{person}{Michael Sedlmair}, {and} \bibinfo{person}{Kuno Kurzhals}.} \bibinfo{year}{2024}\natexlab{}.
\newblock \showarticletitle{Eyes on the {{Task}}: {{Gaze Analysis}} of {{Situated Visualization}} for {{Collaborative Tasks}}}. In \bibinfo{booktitle}{\emph{{{Conference Virtual Reality}} and {{3D User Interfaces}}}}. \bibinfo{publisher}{IEEE}, \bibinfo{pages}{785--795}.
\newblock
\showISBNx{979-8-3503-7403-2}
\urldef\tempurl%
\url{https://doi.org/10.1109/VR58804.2024.00098}
\showDOI{\tempurl}


\bibitem[Pawade et~al\mbox{.}(2018)]%
        {Pawade2018}
\bibfield{author}{\bibinfo{person}{Dipti Pawade}, \bibinfo{person}{Avani Sakhapara}, \bibinfo{person}{Maheshwar Mundhe}, \bibinfo{person}{Aniruddha Kamath}, {and} \bibinfo{person}{Devansh Dave}.} \bibinfo{year}{2018}\natexlab{}.
\newblock \showarticletitle{Augmented {{Reality Based Campus Guide Application}} Using {{Feature Points Object Detection}}}.
\newblock \bibinfo{journal}{\emph{International Journal of Information Technology and Computer Science}} \bibinfo{volume}{10}, \bibinfo{number}{5} (\bibinfo{year}{2018}), \bibinfo{pages}{76--85}.
\newblock
\urldef\tempurl%
\url{https://doi.org/10.5815/ijitcs.2018.05.08}
\showDOI{\tempurl}


\bibitem[Pfeiffer and Renner(2014)]%
        {pfeiffer2014eyesee3d}
\bibfield{author}{\bibinfo{person}{Thies Pfeiffer} {and} \bibinfo{person}{Patrick Renner}.} \bibinfo{year}{2014}\natexlab{}.
\newblock \showarticletitle{{{EyeSee3D}}: A {{Low-Cost Approach}} for {{Analyzing Mobile 3D Eye Tracking Data}} Using {{Computer Vision}} and {{Augmented Reality Technology}}}. In \bibinfo{booktitle}{\emph{{{Symposium}} on {{Eye Tracking Research}} and {{Applications}}}}. \bibinfo{publisher}{ACM}, \bibinfo{pages}{195--202}.
\newblock
\showISBNx{978-1-4503-2751-0}
\urldef\tempurl%
\url{https://doi.org/10.1145/2578153.2578183}
\showDOI{\tempurl}


\bibitem[Pontillo et~al\mbox{.}(2010)]%
        {pontillo2010}
\bibfield{author}{\bibinfo{person}{Daniel~F. Pontillo}, \bibinfo{person}{Thomas~B. Kinsman}, {and} \bibinfo{person}{Jeff~B. Pelz}.} \bibinfo{year}{2010}\natexlab{}.
\newblock \showarticletitle{{{SemantiCode}}: {{Using Content Similarity}} and {{Database-Driven Matching}} to {{Code Wearable Eyetracker Gaze Data}}}. In \bibinfo{booktitle}{\emph{{{Symposium}} on {{Eye Tracking Research}} and {{Applications}}}}. \bibinfo{publisher}{ACM}, \bibinfo{pages}{267--270}.
\newblock
\showISBNx{978-1-60558-994-7}
\urldef\tempurl%
\url{https://doi.org/10.1145/1743666.1743729}
\showDOI{\tempurl}


\bibitem[Raschke et~al\mbox{.}(2014)]%
        {raschke2014}
\bibfield{author}{\bibinfo{person}{Michael Raschke}, \bibinfo{person}{Tanja Blascheck}, \bibinfo{person}{Marianne Richter}, \bibinfo{person}{Tanja Agapkin}, {and} \bibinfo{person}{Thomas Ertl}.} \bibinfo{year}{2014}\natexlab{}.
\newblock \showarticletitle{Visual {{Analysis}} of {{Perceptual}} and {{Cognitive Processes}}}. In \bibinfo{booktitle}{\emph{International {{Conference}} on {{Information Visualization Theory}} and {{Applications}}}}. \bibinfo{publisher}{IEEE}, \bibinfo{pages}{284--291}.
\newblock


\bibitem[Renner and Pfeiffer(2017)]%
        {Renner17}
\bibfield{author}{\bibinfo{person}{Patrick Renner} {and} \bibinfo{person}{Thies Pfeiffer}.} \bibinfo{year}{2017}\natexlab{}.
\newblock \showarticletitle{Attention {{Guiding Techniques}} Using {{Peripheral Vision}} and {{Eye Tracking}} for {{Feedback}} in {{Augmented-Reality-Based Assistance Systems}}}. In \bibinfo{booktitle}{\emph{{{Symposium}} on {{3D User Interfaces}}}}. \bibinfo{publisher}{IEEE}, \bibinfo{pages}{186--194}.
\newblock
\showISBNx{978-1-5090-6717-6}
\urldef\tempurl%
\url{https://doi.org/10.1109/3DUI.2017.7893338}
\showDOI{\tempurl}


\bibitem[Richardson and Dale(2005)]%
        {richardson2005looking}
\bibfield{author}{\bibinfo{person}{Daniel~C Richardson} {and} \bibinfo{person}{Rick Dale}.} \bibinfo{year}{2005}\natexlab{}.
\newblock \showarticletitle{Looking to {{Understand}}: {{The Coupling}} between {{Speakers}}' and {{Listeners}}' {{Eye Movements}} and Its {{Relationship}} to {{Discourse Comprehension}}}.
\newblock \bibinfo{journal}{\emph{Cognitive Science}} \bibinfo{volume}{29}, \bibinfo{number}{6} (\bibinfo{year}{2005}), \bibinfo{pages}{1045--1060}.
\newblock
\urldef\tempurl%
\url{https://doi.org/10.1207/s15516709cog0000_29}
\showDOI{\tempurl}


\bibitem[Richer et~al\mbox{.}(2024)]%
        {Richer:2024:scalability}
\bibfield{author}{\bibinfo{person}{Ga{\"e}lle Richer}, \bibinfo{person}{Alexis Pister}, \bibinfo{person}{Moataz Abdelaal}, \bibinfo{person}{Jean-Daniel Fekete}, \bibinfo{person}{Michael Sedlmair}, {and} \bibinfo{person}{Daniel Weiskopf}.} \bibinfo{year}{2024}\natexlab{}.
\newblock \showarticletitle{Scalability in {{Visualization}}}.
\newblock \bibinfo{journal}{\emph{IEEE Transactions on Visualization and Computer Graphics}} \bibinfo{volume}{30}, \bibinfo{number}{7} (\bibinfo{year}{2024}), \bibinfo{pages}{3314--3330}.
\newblock
\urldef\tempurl%
\url{https://doi.org/10.1109/TVCG.2022.3231230}
\showDOI{\tempurl}


\bibitem[Shepard(1968)]%
        {Shepard1968}
\bibfield{author}{\bibinfo{person}{Donald Shepard}.} \bibinfo{year}{1968}\natexlab{}.
\newblock \showarticletitle{A {{Two-Dimensional Interpolation Function}} for {{Irregularly-Spaced Data}}}. In \bibinfo{booktitle}{\emph{{{National Conference}}}}. \bibinfo{publisher}{ACM}, \bibinfo{pages}{517--524}.
\newblock
\showISBNx{978-1-4503-7486-6}
\urldef\tempurl%
\url{https://doi.org/10.1145/800186.810616}
\showDOI{\tempurl}


\bibitem[Strasburger et~al\mbox{.}(2011)]%
        {strasburger2011peripheral}
\bibfield{author}{\bibinfo{person}{Hans Strasburger}, \bibinfo{person}{Ingo Rentschler}, {and} \bibinfo{person}{Martin J{\"u}ttner}.} \bibinfo{year}{2011}\natexlab{}.
\newblock \showarticletitle{Peripheral {{Vision}} and {{Pattern Recognition}}: {{A Review}}}.
\newblock \bibinfo{journal}{\emph{Journal of Vision}} \bibinfo{volume}{11}, \bibinfo{number}{5} (\bibinfo{year}{2011}), \bibinfo{pages}{13--13}.
\newblock
\urldef\tempurl%
\url{https://doi.org/10.1167/11.5.13}
\showDOI{\tempurl}


\bibitem[Wang et~al\mbox{.}(2021)]%
        {wang2021object}
\bibfield{author}{\bibinfo{person}{Felix Wang}, \bibinfo{person}{Julian Wolf}, \bibinfo{person}{Mazda Farshad}, \bibinfo{person}{Mirko Meboldt}, {and} \bibinfo{person}{Quentin Lohmeyer}.} \bibinfo{year}{2021}\natexlab{}.
\newblock \showarticletitle{Object-{{Gaze Distance}}: {{Quantifying}} near-{{Peripheral Gaze Behavior}} in {{Real-World Applications}}}.
\newblock \bibinfo{journal}{\emph{Journal of Eye Movement Research}} \bibinfo{volume}{14}, \bibinfo{number}{1} (\bibinfo{year}{2021}), \bibinfo{pages}{1--13}.
\newblock
\urldef\tempurl%
\url{https://doi.org/10.16910/jemr.14.1.5}
\showDOI{\tempurl}


\bibitem[Wang et~al\mbox{.}(2023)]%
        {wang2023gaze}
\bibfield{author}{\bibinfo{person}{Felix~Sihan Wang}, \bibinfo{person}{Quentin Lohmeyer}, \bibinfo{person}{Andrew Duchowski}, {and} \bibinfo{person}{Mirko Meboldt}.} \bibinfo{year}{2023}\natexlab{}.
\newblock \showarticletitle{Gaze Is {{More Than}} Just a {{Point}}: {{Rethinking Visual Attention Analysis Using Peripheral Vision-Based Gaze Mapping}}}. In \bibinfo{booktitle}{\emph{{{Symposium}} on {{Eye Tracking Research}} and {{Applications}}}}. \bibinfo{publisher}{ACM}, \bibinfo{pages}{1--7}.
\newblock
\showISBNx{979-8-4007-0150-4}
\urldef\tempurl%
\url{https://doi.org/10.1145/3588015.3589840}
\showDOI{\tempurl}


\bibitem[Wang et~al\mbox{.}(2022)]%
        {Wang2022GazeUncertainty}
\bibfield{author}{\bibinfo{person}{Yao Wang}, \bibinfo{person}{Maurice Koch}, \bibinfo{person}{Mihai B{\^a}ce}, \bibinfo{person}{Daniel Weiskopf}, {and} \bibinfo{person}{Andreas Bulling}.} \bibinfo{year}{2022}\natexlab{}.
\newblock \showarticletitle{Impact of {{Gaze Uncertainty}} on {{AOIs}} in {{Information Visualisations}}}. In \bibinfo{booktitle}{\emph{{{Symposium}} on {{Eye Tracking Research}} and {{Applications}}}}. \bibinfo{publisher}{ACM}, Article \bibinfo{articleno}{60}, \bibinfo{numpages}{6}~pages.
\newblock
\urldef\tempurl%
\url{https://doi.org/10.1145/3517031.3531166}
\showDOI{\tempurl}


\bibitem[Weibel et~al\mbox{.}(2012)]%
        {weibel2012let}
\bibfield{author}{\bibinfo{person}{Nadir Weibel}, \bibinfo{person}{Adam Fouse}, \bibinfo{person}{Colleen Emmenegger}, \bibinfo{person}{Sara Kimmich}, {and} \bibinfo{person}{Edwin Hutchins}.} \bibinfo{year}{2012}\natexlab{}.
\newblock \showarticletitle{Let's {{Look}} at the {{Cockpit}}: {{Exploring Mobile Eye-Tracking}} for {{Observational Research}} on the {{Flight Deck}}}. In \bibinfo{booktitle}{\emph{{{Symposium}} on {{Eye Tracking Research}} and {{Applications}}}}. \bibinfo{publisher}{ACM}, \bibinfo{pages}{107--114}.
\newblock
\showISBNx{978-1-4503-1221-9}
\urldef\tempurl%
\url{https://doi.org/10.1145/2168556.2168573}
\showDOI{\tempurl}


\bibitem[Weiskopf(2022)]%
        {weiskopf2022uncertainty}
\bibfield{author}{\bibinfo{person}{Daniel Weiskopf}.} \bibinfo{year}{2022}\natexlab{}.
\newblock \showarticletitle{Uncertainty {{Visualization}}: {{Concepts}}, {{Methods}}, and {{Applications}} in {{Biological Data Visualization}}}.
\newblock \bibinfo{journal}{\emph{Frontiers in Bioinformatics}}  \bibinfo{volume}{2} (\bibinfo{year}{2022}), \bibinfo{pages}{793819}.
\newblock
\urldef\tempurl%
\url{https://doi.org/10.3389/fbinf.2022.793819}
\showDOI{\tempurl}


\bibitem[Wolf et~al\mbox{.}(2018)]%
        {wolf2018automating}
\bibfield{author}{\bibinfo{person}{Julian Wolf}, \bibinfo{person}{Stephan Hess}, \bibinfo{person}{David Bachmann}, \bibinfo{person}{Quentin Lohmeyer}, {and} \bibinfo{person}{Mirko Meboldt}.} \bibinfo{year}{2018}\natexlab{}.
\newblock \showarticletitle{Automating {{Areas}} of {{Interest Analysis}} in {{Mobile Eye Tracking Experiments Based}} on {{Machine Learning}}}.
\newblock \bibinfo{journal}{\emph{Journal of Eye Movement Research}} \bibinfo{volume}{11}, \bibinfo{number}{6} (\bibinfo{year}{2018}), \bibinfo{pages}{1--11}.
\newblock
\urldef\tempurl%
\url{https://doi.org/10.16910/jemr.11.6.6}
\showDOI{\tempurl}


\bibitem[Yang et~al\mbox{.}(2020)]%
        {yang2020comparison}
\bibfield{author}{\bibinfo{person}{Chia-Kai Yang}, \bibinfo{person}{Tanja Blascheck}, {and} \bibinfo{person}{Chat Wacharamanotham}.} \bibinfo{year}{2020}\natexlab{}.
\newblock \showarticletitle{A {{Comparison}} of a {{Transition-Based}} and a {{Sequence-Based Analysis}} of {{AOI Transition Sequences}}}. In \bibinfo{booktitle}{\emph{{{Symposium}} on {{Eye Tracking Research}} and {{Applications}}}}. \bibinfo{publisher}{ACM}, \bibinfo{pages}{1--5}.
\newblock
\showISBNx{978-1-4503-7134-6}
\urldef\tempurl%
\url{https://doi.org/10.1145/3379156.3391349}
\showDOI{\tempurl}


\bibitem[Yang and Wacharamanotham(2018)]%
        {Yang2018alpscarfs}
\bibfield{author}{\bibinfo{person}{Chia-Kai Yang} {and} \bibinfo{person}{Chat Wacharamanotham}.} \bibinfo{year}{2018}\natexlab{}.
\newblock \showarticletitle{Alpscarf: {{Augmenting Scarf Plots}} for {{Exploring Temporal Gaze Patterns}}}. In \bibinfo{booktitle}{\emph{Extended {{Abstracts}} of the {{Conference}} on {{Human Factors}} in {{Computing Systems}}}}. \bibinfo{publisher}{ACM}, \bibinfo{pages}{1--6}.
\newblock
\showISBNx{978-1-4503-5621-3}
\urldef\tempurl%
\url{https://doi.org/10.1145/3170427.3188490}
\showDOI{\tempurl}


\end{thebibliography}


\clearpage
\appendix
\section*{Supplemental Material}
\section{Experiment: Eye-Tracking in AR}
With an example experiment, we demonstrate the usefulness of uncertainty-aware scarf plots.
The experiment recordings comprise multiple sources of uncertainty to showcase the suggested techniques in comparison with standard scarf plots as a baseline.

\subsection{Setup}

The experiment was set up as a laboratory experiment under controlled environmental conditions. We used a HoloLens~2 to record gaze in an AR scene. For this, we prepared a table with physical and virtual objects (see \autoref{fig:experiment}). 
For recording gaze data, we used the ARETT\footnote{\url{https://github.com/AR-Eye-Tracking-Toolkit/ARETT}} package by \citet{Kapp2021ARETT}. For object detection, we utilized the Unity project \textit{hl2\_detection}\footnote{\url{https://github.com/kolaszko/hl2_detection}} by \citet{lysakowski2023YOLO}. They proposed a software architecture that allows running the \textit{YOLOv8}~\cite{yolov8_ultralytics} network directly on the HoloLens~2, providing real-time object detection. 

For our application, we imported both projects into our Unity\footnote{\url{https://unity.com/de}} project. Since ARETT only reports intersections of the gaze ray with the first hit of a virtual object, we replaced the \textit{Physics.Raycast()} function with the \textit{Physics.RaycastNonAlloc()} function, which can be used to get a specified number of intersections along the defined ray. We also modified the \textit{DataLogger} script in ARETT to save the virtual objects intersected by the gaze ray and the corresponding gaze points in a CSV file.  
The intersection test with physical objects was implemented in the \textit{Detection} script of the \textit{hl2\_detection} project. 
To ensure object detection even if a virtual object is in front of the object, we disabled the \textit{enablehologram} parameter within the camera parameters of the \textit{Detection} script. 

\subsection{Scene Composition}
We designed a composition of multiple virtual and physical objects that were inspected in specific orders and from different angles to evoke specific patterns of uncertainty in the recordings.
\paragraph{\textbf{Two bottles (BB)}} 
For the first task, we displayed the scenario shown in \autoref{fig:2bottles}. The participants were asked to view two bottles placed diagonally on the table. We instructed the participants to view the bottles first at the intersection region of the bounding boxes. Afterward, they shifted their gaze to the bottle in the front and then to the one at the back. In this task, we targeted the ambiguity developed when looking at an intersection area of two objects in different depths. This corresponds to the uncertainty of order. Further, we chose the same object to demonstrate how the uncertainty-aware scarf plot will handle two objects with the same label for the AOI. 

\paragraph{\textbf{Bottle between virtual plant and virtual bust (VP, B \& VB)}} 
We placed a bottle between two virtual objects, as shown in \autoref{fig:virutalbottle}. The task required looking at the objects at different depths sequentially, first from the closest to the farthest and back. The participants began by focusing on the virtual plant. With this task, we aimed to produce the uncertainties occurring when a participant can view another object through a virtual one, due to transparency. 

Since our implementation also considers the intersection with objects at different depths, we can provide a visualization that deals with the gaze ray intersection of multiple objects at different depths, regardless of whether they are virtual or physical.

\paragraph{\textbf{Virtual plant and bust (VP \& VB)}}
For the stimuli shown in \autoref{fig:virtual}, the participants had to look at two virtual objects at different depths, without an additional physical object. The task was to shift the gaze between the two objects by starting first at the virtual plant. In this case, we expected uncertainties regarding order and position, similar to the task \textit{BB}. However, the virtual objects did not compensate for positional offset as the physical objects would. Therefore, gaze rays along the edge of the virtual plant will more often lead to one intersection on the virtual bust, and therefore, the depth scarf plot will look more similar to the standard scarf plot in these cases. However, the NN scarf plot will still consider both objects when the gaze ray is in proximity to both virtual objects, accounting for the uncertainty in position.

\paragraph{\textbf{Virtual plant, cup, and virtual bust (VP, C \& VB)}}
The last task involved again two virtual objects and one physical object, as depicted in \autoref{fig:virtualandcup}. We placed a physical cup next to the virtual plant at approximately the same depth. The participants were asked to look at the plant, then the cup, and afterward the bust, and vice versa. In this case, we do not cover the uncertainty caused by the order. Therefore, the standard scarf plot and depth scarf plot should be similar. In this case, we expect the NN scarf plot to provide insightful visualization, as the spatial proximity of the three objects could introduce uncertainties in terms of position. 

\section{Results -- Scarf Plots for 4 Different Tasks}
\begin{figure*}[!h]
  \centering
      \includegraphics[width=1.0\textwidth]{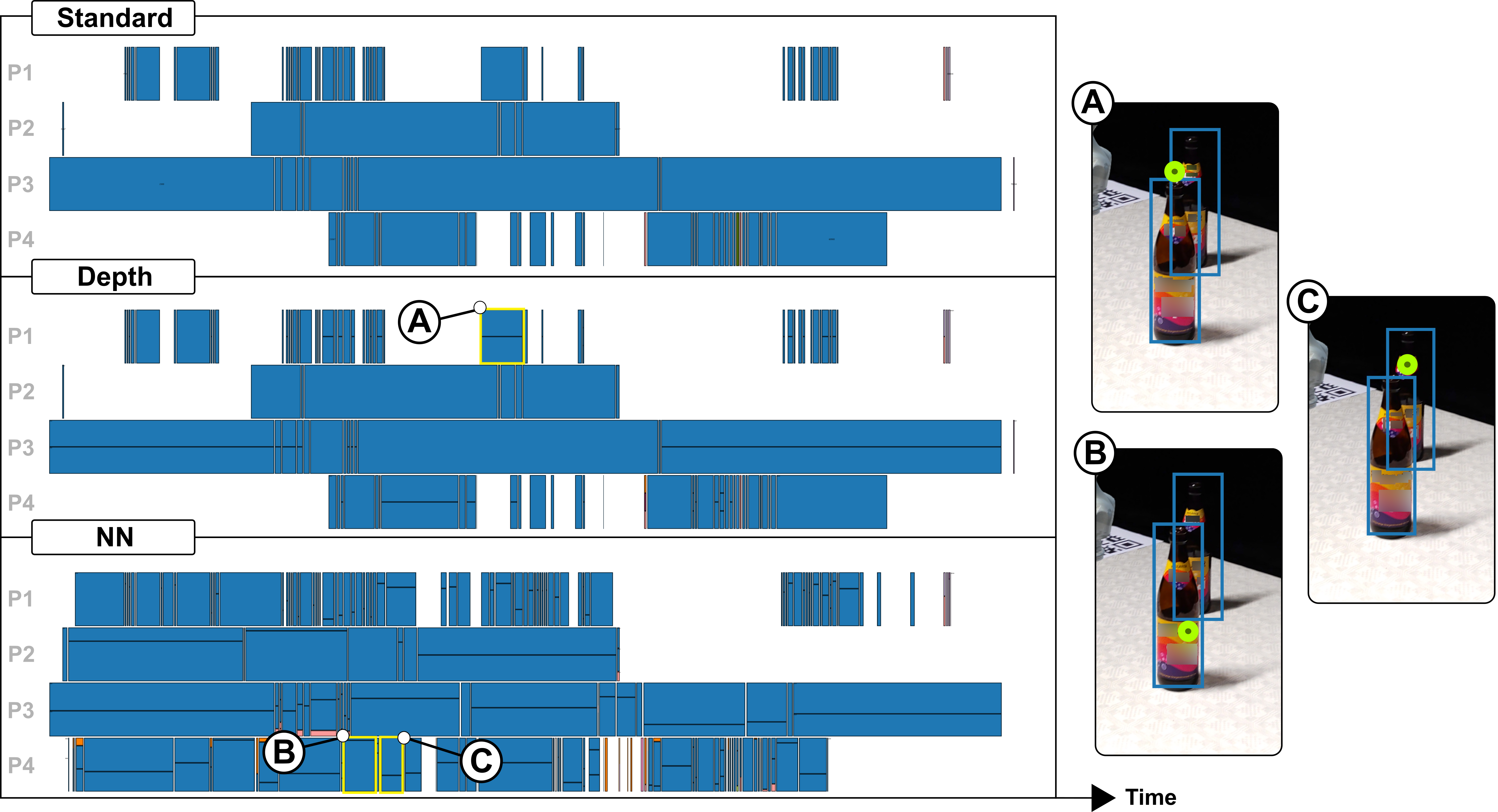}
      \caption{The scarf plots for the task \textit{BB}. Each row (P1--P4) in the plots belongs to the data set of a participant. The plots mainly visualize the AOIs of the bottles; therefore, all plots primarily contain blue boxes for these AOIs. (Top) The standard scarf plot makes little sense here, showing only the first ray intersection.
      (Center) In contrast, the depth plot shows  multiple segments divided into two, indicating that the participants' gaze ray intersected with the bounding box of both bottles (see (A)). (Bottom)~With the NN scarf plot, we can understand when the participants were switching their gaze to another bottle (see (B) and (C)). The white spaces in the plots indicate time spans in which no gaze data was available or no hit was detected on the gaze ray or close to it.}\label{fig:BottlesScarf}
\end{figure*}

\begin{figure*}[!h]
    \centering
    \includegraphics[width=1.0\textwidth]{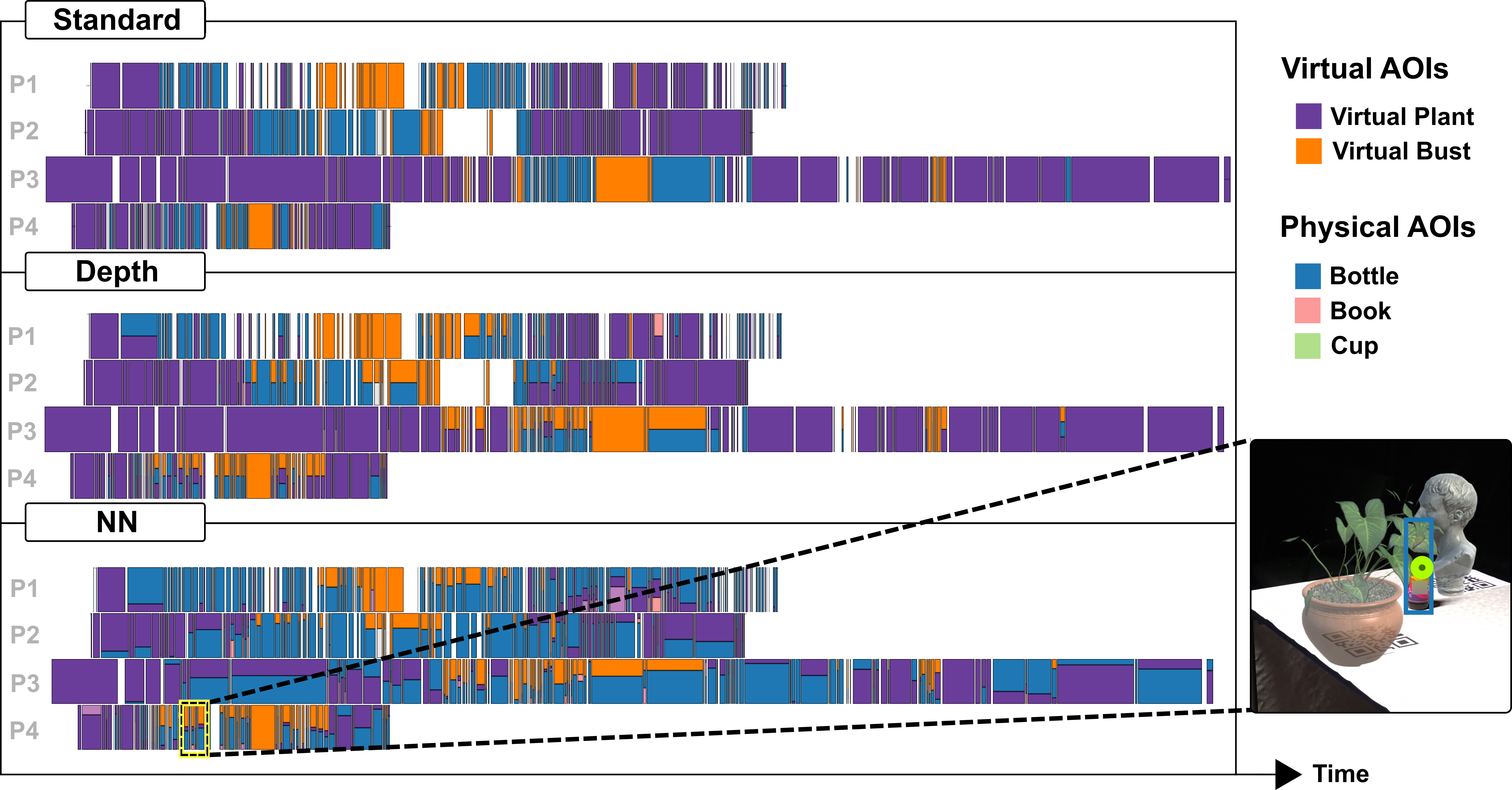}
    \caption{The scarf plots for the task \textit{VP, B \& VB}. Each row (P1--P4) in the plots belongs to the data set of a participant. (Top) The standard scarf plot shows the main AOIs present in the stimulus. 
    (Center) The depth scarf plot provides an interpretation of the depth of the different AOIs over time. 
    (Bottom) The NN plot is similar to the depth plot but provides insight into which AOI the participant is more likely to have looked at during specific time steps, accounting for positional uncertainties. The example shows segments where the gaze point lies on the bottle placed between the virtual plant and the virtual bust. Since the virtual objects are close to the bottle, the segment contains the color of all three AOIs as sub-segments, with the bottle and the virtual bust taking up the majority of space, since the gaze ray is the closest to those two objects. The bottle sub-segment is on the bottom since the object detection algorithm sometimes detects the bottle in front of the virtual plant. The white spaces in the plots indicate time spans in which no gaze data was available or no hit was detected on the gaze ray or close to it. \scriptsize (Virtual bust from guptaarnish: \textcopyright~ Sketchfab Standard License: \url{https://sketchfab.com/3d-models/marble-bust-01-4k-c9b1839068094e40bc941b047ca19a85})
    }
    \label{fig:virtualBottleScarf}
\end{figure*}

\begin{figure*}[!h]
    \centering
    \includegraphics[width = 1.0\textwidth]{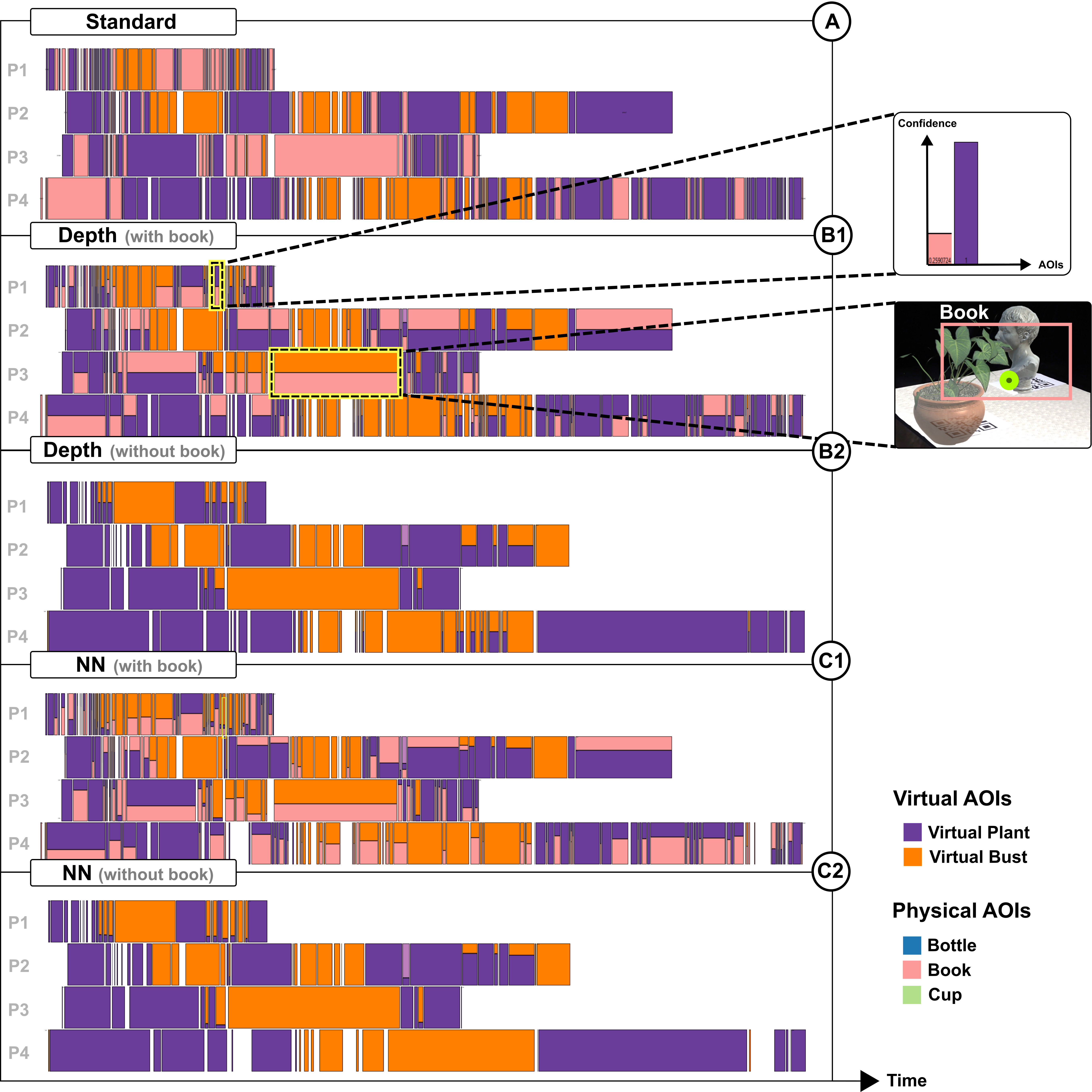}
    \caption{The scarf plot for the task \textit{VP \& VB}. Each row (P1--P4) in the plots belongs to the data set of a participant. The scarf plots (A), (B1), (C1) show that a non-existent AOI (Book) was detected. By clicking on one of the segments, the analysts can view the bar chart showing the percentage of confidence for the detected AOI. Due to the misclassification, the standard scarf plot (A) provides wrong information regarding the gaze pattern. Since the depth scarf plot (B1) shows gaze on different depths, the falsely detected AOI can be ignored, when viewing the scarf plot. (B2) shows a version where the book was filtered out. The NN plot (C1) is not useful in this situation, since some areas are covered by the color corresponding to the non-existent AOI. Here, it is important to filter out the wrong annotation. (C2) allows us to interpret when the participant was focusing on which virtual object and when the gaze point was in between both objects. The white spaces in the plots indicate time spans in which no gaze data was available or no hit was detected on the gaze ray or close to it. \scriptsize (Virtual bust from guptaarnish: \textcopyright~ Sketchfab Standard License: \url{https://sketchfab.com/3d-models/marble-bust-01-4k-c9b1839068094e40bc941b047ca19a85})}
    \label{fig:virtualDepthScarf}
\end{figure*}

\begin{figure*}[!h]
    \centering
    \includegraphics[width = 1.0\textwidth]{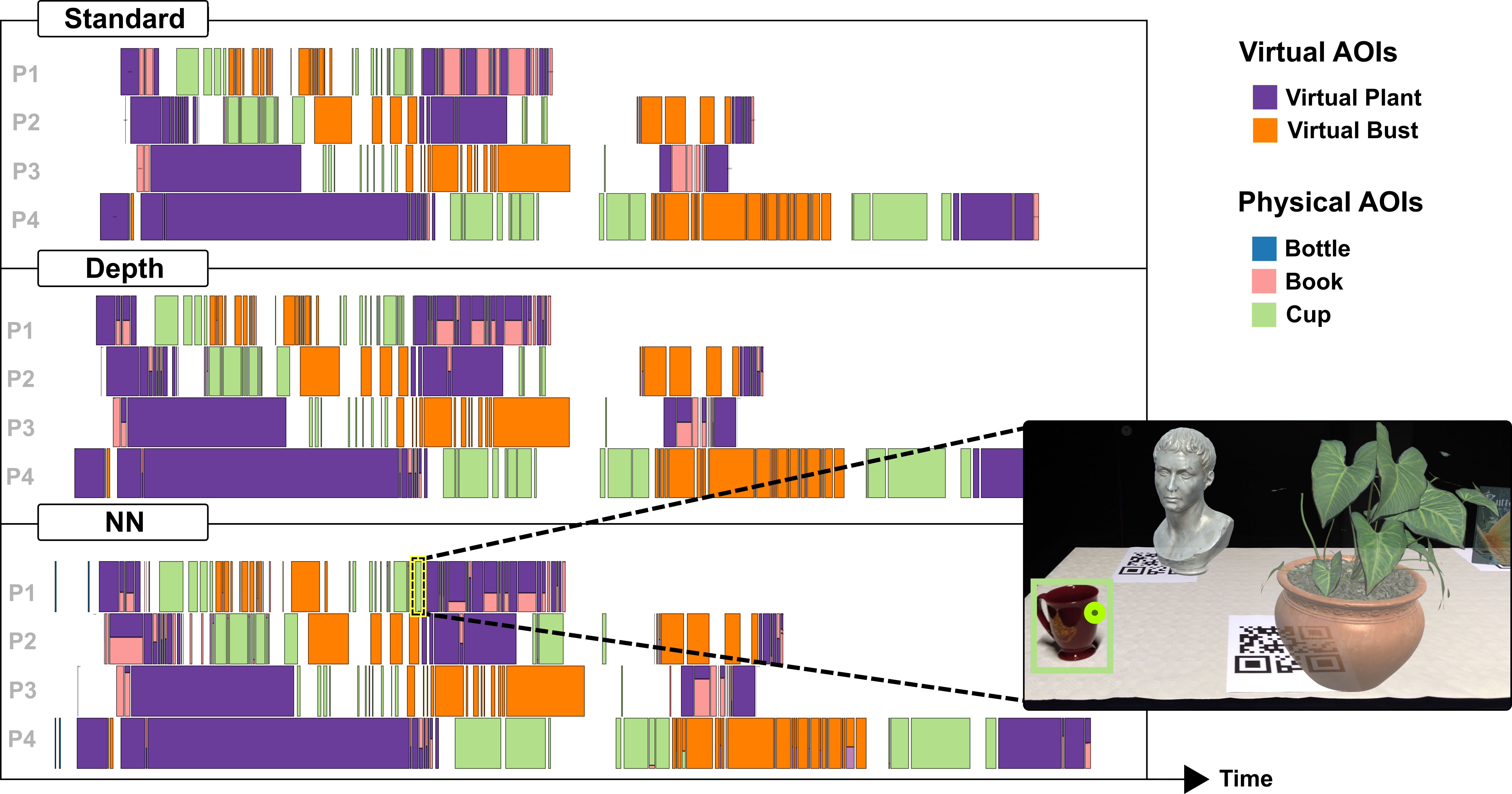}
    \caption{The scarf plot for task \textit{VP, C \& VB}. Each row (P1--P4) in the plots belongs to the data set of a participant. In this scenario, (Top) the standard scarf plot, (Center) depth scarf plot, and (Bottom) the NN plot are quite similar, which shows that this kind of setup is less prone to uncertainty occurring through depth and position. The white spaces in the plots indicate time spans in which no gaze data was available or no hit was detected on the gaze ray or close to it. \scriptsize (Virtual bust from guptaarnish: \textcopyright~ Sketchfab Standard License: \url{https://sketchfab.com/3d-models/marble-bust-01-4k-c9b1839068094e40bc941b047ca19a85})}
    \label{fig:virtualDepthCupScarf}
\end{figure*}

We compared the suggested extensions \textit{depth} and \textit{NN} with the standard scarf plots for the four tasks, as discussed in the paper. 
\autoref{fig:BottlesScarf}, \autoref{fig:virtualBottleScarf}, \autoref{fig:virtualDepthScarf}, and \autoref{fig:virtualDepthCupScarf} present the different scarf plots of all participants per task. Each row (P1--P4) in the different scarf plots corresponds to the recorded data set of one participant. The white spaces within the scarf plots correspond to time spans in which there is no gaze available or no hit with an AOI. 

\section{Discussion -- Sources of Uncertainty not Covered}
Our experiment demonstrates that uncertainty-aware scarf plots can show the most common potential issues in eye-tracking data occurring mainly through positional offset, depth, and automatic detection of AOIs. However, there are additional cases, some related to stimuli and some to specific hardware setups, that we did not cover with the presented approach.

\paragraph{Gaze Confidence}
Some eye-tracking devices provide information about the quality of the detected gaze samples \cite{holmqvist2012eye}. To this point, we did not include this aspect due to its limited availability with some devices.

\paragraph{Recording Quality}
Another source of potential inaccuracies is synchronization issues with some devices where gaze and video are recorded separately via wireless networks.
For instance, frame drops in recorded video streams are often hard to register and can cause issues when linking gaze data back to the stimulus.
In general, information about the quality of the stimulus video is another aspect worth investigating for future expansions of the scarf plot visualization. Frame drops could be displayed by empty segments in the scarf plot. 
If the respective gaze data is available and no frames are available, this could be displayed with a binary encoding, e.g., by adding tick marks to affected segments.
\paragraph{Misclassification of Dynamic AOIs}
As our examples have shown, there can be some misclassified objects in the scene. We provide information regarding the confidence of prediction through a bar chart, but there is no visualization encoding the correctness of labeling. When the stimuli are known beforehand, labels that do not belong to the scene can be easily detected in the visualization and filtered out. However, in dynamic scenes, the analysts are not aware of all AOIs within the scene, and therefore, it gets more challenging to filter out wrong predictions during the pre-processing of data.  

\paragraph{Calibration}
Calibration plays an important role in achieving a correct mapping of gaze onto the stimuli. However, head-worn eye-tracking devices can cause loss of calibration, due to unwanted movement of the headsets, leading to incorrect mapping of gaze to~stimuli.
\paragraph{Sampling Rate}
The sampling rate of eye-tracking devices is crucial for accurate tracking and good precision. A low sampling rate can result in shifts of gaze samples in time, leading to gaze samples having an offset in time~\cite{Andersson2010SamplingFreq}.

\section{Additional Supplemental}
\begin{figure*}[htbp]
    \centering
    \includegraphics[width=0.9\textwidth]{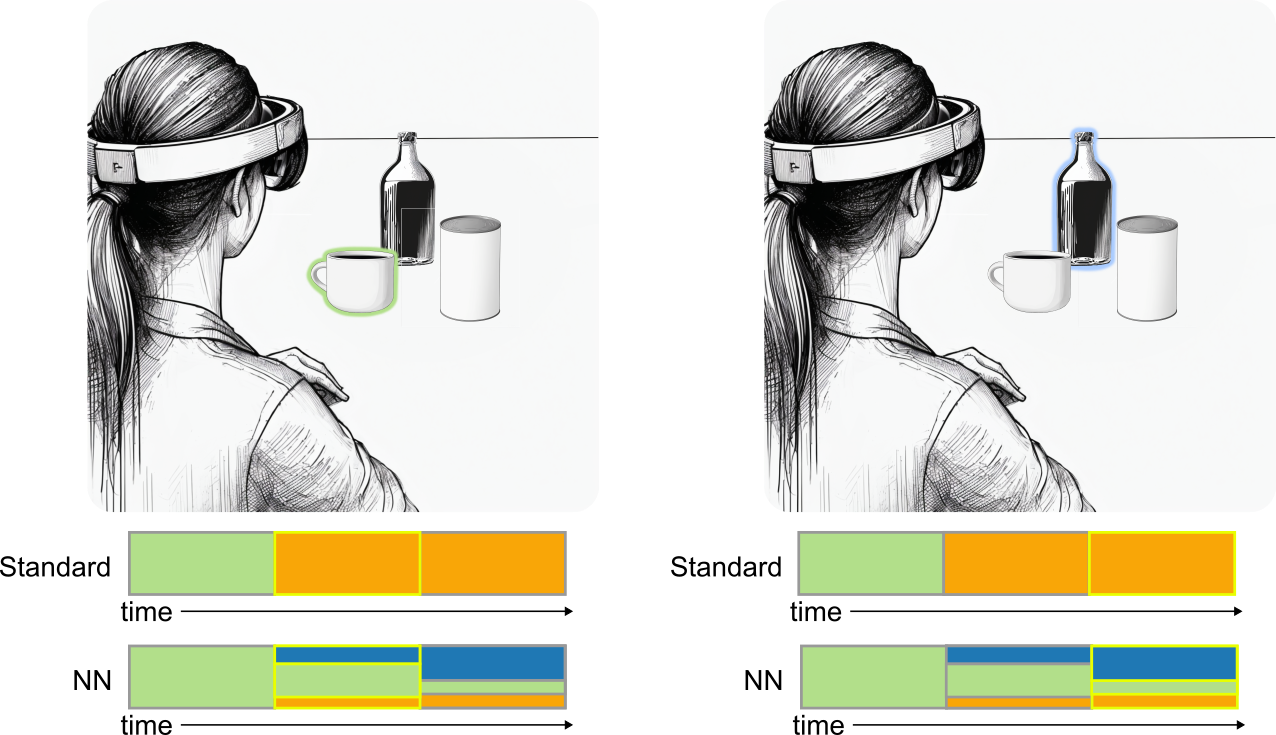}
    \caption{
    Highlighting objects can help to guide the attention of a user in an environment. Here, the attention is first guided toward the cup, then the bottle. However, since the table is crowded by multiple objects, the \textit{standard scarf plot}  leads to wrong gaze patterns. The \textit{NN scarf plot}, however, shows on which object the gaze ray was the closest, allowing the detection of the expected pattern.  (Image generated with Stable Diffusion)
    }
    \label{fig:visualsearchtask}
\end{figure*}

\paragraph{Illustration of visual search task}
\autoref{fig:visualsearchtask} illustrates an attention guidance technique in which the object of interest is highlighted to direct the user's focus. However, when analyzing the user's gaze pattern, the \textit{standard scarf plot} may reveal unexpected results. This is because the highlighted object is positioned close to other objects, and inaccuracies in the eye-tracking device may cause the gaze ray to intersect with a neighboring object, even when the user is actually looking at the highlighted one. For instance, in ~\autoref{fig:visualsearchtask}, the highlighted cup is near both the bottle and the tin. As a result, the \textit{standard scarf plot} might indicate an intersection with the bottle or the tin. In this example, the \textit{standard scarf plot} shows an intersection with the tin. In contrast, the \textit{NN scarf plot} represents all the objects in proximity to the gaze ray, while the object that is the closest to the gaze ray is more prominent in the scarf plot, revealing the expected gaze pattern. In ~\autoref{fig:visualsearchtask}, we can see how the area of the cup sub-segment is larger in the second segment of the scarf plot, and then the area of the bottle sub-segment increases in the third segment. We highlighted the segment (yellow) that corresponds to the current state of the stimuli.
\end{document}